\documentclass[11pt,psfig,a4]{article}
\usepackage[hmarginratio=2:3,top=32mm,left=20mm,columnsep=17pt]{geometry}

\usepackage{lineno,array}
\usepackage{titling}
\usepackage{color}
\usepackage{tabularx}
\usepackage{graphicx,epstopdf}
\usepackage[dvipsnames]{xcolor}
\usepackage{amsmath,amsthm,thmtools}
\usepackage{amssymb}
\usepackage{amsfonts}
\usepackage{moreverb}
\usepackage{dsfont}
\usepackage{tipa}
\usepackage{upgreek}
\usepackage{bm}
\usepackage{multirow}
\usepackage{soul}
\usepackage[normalem]{ulem} %to strike the words
\usepackage{ textcomp }
\usepackage{relsize} % e.g. used for \mathsmaller
\usepackage{xfrac}
\usepackage{lastpage} % for the number of the last page in the document
\usepackage{fancyhdr}
\usepackage{wrapfig}
\pdfoutput=1
\usepackage[colorlinks]{hyperref}

\usepackage{xifthen}% provides \isempty test

\usepackage[backend=bibtex,giveninits=true,url=false,doi=true,eprint=true,isbn=false,
backref,backrefstyle=none,maxbibnames=99]{biblatex}
\DefineBibliographyStrings{english}{%
  backrefpage = {Cited on p\adddot},%
  backrefpages = {Cited on pp\adddot}%
}

\bibliography{library}
\renewbibmacro{in:}{%
  \ifboolexpr{%
     test {\ifentrytype{article}}%
  }{}{\printtext{\bibstring{in}\intitlepunct}}%
}
\definecolor{mygray}{rgb}{0.0,0.0,0.0}
\declaretheoremstyle[
  headfont=\color{mygray}\normalfont\bfseries,
  bodyfont=\color{mygray}\normalfont\footnotesize,
]{colored}
\declaretheorem[
  style=colored,
  name=Remark,
]{remark}
\makeatletter

\newcommand{\pd}{\partial}
\newcommand{\rmd}{{\rm d}}

\newcommand{\Dist}{{\bm{A}}}						% Distortion matrix, i.e. inverse F
\newcommand{\calE}{\mathcal{E}}						%
\newcommand{\Burg}{{\bm{B}}}
\newcommand{\Durg}{{\bm{D}}}
\renewcommand{\Re}{\text{Re}}

\newcommand{\Cauchystr}[2]{\Sigma^{#1}_{\phantom{#1}#2}}
\newcommand{\MomFlux}[2]{T^{#1}_{\phantom{#1}#2}}

\renewcommand{\AA}{\bm{A}}
\newcommand{\GG}{\bm{G}}

\newcommand{\vv}{\bm{v}}
\newcommand{\mm}{\bm{m}}
\newcommand{\MM}{\bm{M}}

\newcommand{\sA}{\mathsmaller A}
\newcommand{\sB}{\mathsmaller B}
\newcommand{\sC}{\mathsmaller C}

\newcommand{\myxi}[1]{ \xi^{\mathsmaller #1} }	% Distortion field A

\newcommand{\veps}{\varepsilon}
\newcommand{\transpose}{{\rm {\mathsmaller T}}}

\newcommand{\timeunit}{\text{T}}
\newcommand{\length}{\text{L}}
\newcommand{\mass}{\text{M}}
\newcommand{\Vel}{\text{V}}

\newcommand{\dof}{DoF }

\newcommand{\triad}[2]{ A^{#1}_{\phantom{#1}#2} }	% Distortion field A
\newcommand{\itriad}[2]{ E^{#1}_{\phantom{#1}#2} }		% Inverse triad
\newcommand{\itetrad}[2]{ E^{#1}_{\ #2} }		% Inverse triad
\newcommand{\dualtors}[1]{{T^*}^{\hspace{-0.2mm}#1}}
\newcommand{\dist}[2]{ A^{#1}_{\phantom{#1}#2} }	% Component of the distortion field A

\newcommand{\defgrad}[2]{ F^{#1}_{\ \mathsmaller#2} }		% Def gradient
\newcommand{\idefgrad}[2]{ F^{\mathsmaller#1}_{\ #2} }		% Def gradient
\newcommand{\burg}[2]{ B^{{#1}{#2}} }	% Eulerian components of the B field
\newcommand{\durg}[2]{ D_{#1}^{\phantom{#1}#2} }	% Eulerian components of the D field
\newcommand{\burgL}[2]{ B^{{#1}{#2}} }	% Eulerian components of the B field
\newcommand{\durgL}[2]{ D^{#1}_{\phantom{#1}#2} }	% Eulerian components of the D field
\newcommand{\vel}[1]{v^{#1}}
\newcommand{\LeviCivitaUp}[1]{\varepsilon^{#1}}
\newcommand{\kronecker}[2]{\delta^{#1}_{\phantom{#1}#2}}

\newcommand{\tors}[2]{ T^{#1}_{\ #2} }	% Torsion
\newcommand{\W}[2]{ W^{#1}_{\ #2} }	% Weitzenbock connection
\newcommand{\motion}[1]{\chi^{{#1}}}
\newcommand{\motionopt}[1][]{%
  \ifthenelse{\isempty{#1}}%
    {\chi}% if #1 is empty
    {\chi^{#1}}% if #1 is not empty
					    }

\newcommand{\Matter}{M^{t}}		% Material manifold
\newcommand{\MatterO}{M^{0}} 	% material manifold at t = 0
\newcommand{\MatterR}{M^{t,{\rm rel}}}	% Relaxed material manifold
\newcommand{\TM}{T_xM^{t}}		% Material manifold
\newcommand{\TMlr}{T^{\rm rel}_xM^t}	% Relaxed material manifold
\newcommand{\TMgr}{T_{\xi(x)}M^0} 	% material manifold at t = 0
\newcommand{\Rot}[2]{R^{#1'}_{\ #2}}
\newcommand{\Qot}[2]{Q^{#1}_{\ #2'}}
\newcommand{\ee}{\bm{e}}
\newcommand{\csh}{c_\text{\tiny sh}}	%	shear sound speed
\newcommand{\csp}{c_\text{\tiny sp}}	% 	a sound speed related to the spin

\newcommand{\Weitz}{Weitzenb\"ock\ } 

\newcommand*\samethanks[1][\value{footnote}]{\footnotemark[#1]}

\usepackage{xargs}                      % Use more than one optional parameter in a new commands
\usepackage[colorinlistoftodos,prependcaption,textsize=tiny]{todonotes}
\newcommandx{\noteOne}[2][1=]{\todo[linecolor=red,backgroundcolor=red!25,bordercolor=red,#1]{#2}}
\newcommandx{\noteTwo}[2][1=]{\todo[linecolor=blue,backgroundcolor=blue!25,bordercolor=blue,#1]{#2}}
\newcommandx{\noteMy}[2][1=]{\todo[linecolor=ForestGreen,backgroundcolor=YellowGreen!25,bordercolor=ForestGreen,#1]{#2}}

 \usepackage{empheq}
 \newlength\mytemplen
 \newsavebox\mytempbox
 \makeatletter
 \definecolor{myblue}{rgb}{.81, .88, 1}
 \newcommand\mybluebox{%
     \@ifnextchar[%]
        {\@mybluebox}%
        {\@mybluebox[0pt]}}
 \def\@mybluebox[#1]{%
     \@ifnextchar[%]
        {\@@mybluebox[#1]}%
        {\@@mybluebox[#1][0pt]}}
 \def\@@mybluebox[#1][#2]#3{
     \sbox\mytempbox{#3}%
     \mytemplen\ht\mytempbox
     \advance\mytemplen #1\relax
     \ht\mytempbox\mytemplen
     \mytemplen\dp\mytempbox
     \advance\mytemplen #2\relax
     \dp\mytempbox\mytemplen
     \colorbox{myblue}{\hspace{1em}\usebox{\mytempbox}\hspace{1em}}}
 \makeatother

 \makeatletter
\definecolor{cream}{rgb}{1.0, 0.99, 0.82}
 \newcommand\mycreambox{%
     \@ifnextchar[%]
        {\@mycreambox}%
        {\@mycreambox[0pt]}}
 \def\@mycreambox[#1]{%
     \@ifnextchar[%]
        {\@@mycreambox[#1]}%
        {\@@mycreambox[#1][0pt]}}
 \def\@@mycreambox[#1][#2]#3{
     \sbox\mytempbox{#3}%
     \mytemplen\ht\mytempbox
     \advance\mytemplen #1\relax
     \ht\mytempbox\mytemplen
     \mytemplen\dp\mytempbox
     \advance\mytemplen #2\relax
     \dp\mytempbox\mytemplen
     \colorbox{cream}{\hspace{1em}\usebox{\mytempbox}\hspace{1em}}}
 \makeatother

\hypersetup{
    citecolor=RoyalBlue,
    linkcolor=RedViolet,
	urlcolor=NavyBlue
 }
\fancypagestyle{firstpagestyle}
{
	\fancyhead[C]{} 
	\fancyhf{}
	\fancyfoot[L]{\footnotesize\it Preprint submitted to CMAT}
}
\title{\vspace{-15mm}\fontsize{16pt}{10pt}\bf\selectfont{Continuum mechanics with torsion}}

\author{
\textsc{Ilya Peshkov,}\thanks{Institut de Math\'{e}matiques de Toulouse, 
(Toulouse, France), 
\href{mailto:peshenator@gmail.com}{e-mail}, 
}$\ ^,$\samethanks[2]
\quad 
\textsc{Evgeniy Romenski,}\thanks{Sobolev 
Institute of Mathematics (Novosibirsk, 
Russia), \href{mailto:evrom@math.nsc.ru}{e-mail}}$\ ^, $\thanks{Novosibirsk State University 
(Novosibirsk, Russia).}$\ ^,$\samethanks[4]\quad
\textsc{Michael Dumbser,}\thanks{Department of Civil, Environmental and Mechanical Engineering, 
University of Trento,  (Trento, 
Italy), \href{mailto:michael.dumbser@unitn.it}{e-mail}, 
}
}
\thanksmarkseries{arabic}

\begin{document}
\maketitle
\thispagestyle{empty}

% PARAGRAPH OPTIONS:
\setlength\parindent{0pt} % sets indent to zero
\setlength{\parskip}{5pt} % changes vertical space between paragraphs
% PARAGRAPH OPTIONS.

% ABSTRACT
\begin{abstract}
\noindent
This paper is an attempt to introduce methods and concepts of the Riemann-Cartan geometry 
largely used in such physical theories as general relativity, gauge theories, solid dynamics, etc. 
to fluid dynamics in general and to studying and modeling turbulence in particular. Thus, in 
order to account for the rotational degrees of freedom of the irregular dynamics of small scale 
vortexes, we further generalize our unified first-order hyperbolic formulation of continuum fluid 
and solid mechanics which treats the flowing medium as a Riemann-Cartan manifold with zero 
curvature but non-vanishing torsion. We associate the rotational degrees of freedom of the 
main field of our theory, the distortion field, to the dynamics of microscopic (unresolved) 
vortexes. The distortion field characterizes the deformation and rotation of the material elements 
and can be viewed as anholonomic basis triad with non-vanishing torsion. The torsion tensor is then 
used to characterize distortion's spin and is treated as an independent field with its own time 
evolution equation. This new governing equation has essentially the structure of the non-linear 
electrodynamics in a moving medium and can be viewed as a Yang-Mills-type gauge 
theory. The system is closed by providing an example  of the total energy potential. The extended 
system describes not only irreversible dynamics (which raises the entropy) due to the viscosity or 
plasticity effect but it also has dispersive features which are due to the reversible energy 
exchange (which conserves the entropy) between micro and macro scales. Both the irreversible and 
dispersive processes are represented by relaxation-type algebraic source terms so that the overall 
system remains first-order hyperbolic. The turbulent state is then treated as an excitation of the 
equilibrium laminar state due to the non-linear interplay between dissipation and dispersion.
\end{abstract}

%\linenumbers
\thispagestyle{firstpagestyle}

\section{Introduction}\label{sec.intro}

A unified formulation for continuum fluid and solid mechanics was proposed recently 
in~\cite{HPR2016,DPRZ2016,DPRZ2017,HYP2016}.
In a one system of governing partial differential equations (PDEs), it is able to describe the 
irreversible dynamics of viscous
fluids, either Newtonian~\cite{DPRZ2016} or non-Newtonian~\cite{Jackson2018}, and 
elastoplastic~\cite{GodRom2003,Hyper-Hypo2018,HYP2016,GodPesh2010,BartonRom2010} 
solids as well as the reversible dynamics of inviscid fluids and elastic 
solids. Such 
a formulation relies on the
theory of Symmetric Hyperbolic Thermodynamically Compatible 
equations (SHTC)~\cite{SHTC-GENERIC-CMAT,GodRom2003,Rom1998,Godunov1996,God1961}, which implies 
that the governing 
PDEs are 
first-order
hyperbolic equations. Also, the model is grounded on geometrical principles and Hamiltonian's 
principle of stationary action~\cite{SHTC-GENERIC-CMAT} which are known to be valid in diverse 
physical theories. For instance, this opens attractive 
new possibilities for designing of models in a \textit{first-principle-based} 
fashion in contrast to the classical/modern phenomenology prevailing Navier-Stokes-based models in 
fluid dynamics in general and turbulence modeling in particular. 

The first goal of this paper is to demonstrate that the original unified 
formulation of continuum mechanics
can be generalized in a thermodynamically consistent way in order
to take into account the rotational degrees of freedom (DoF) of the microstructure (either 
emerging 
like in turbulence or plasticity or preset like in microstructured solids) encoded in the key 
field 
of the 
theory, the 
\textit{distortion field}. Our second goal is to introduce powerful and universally valid methods 
of differential geometry to fluid dynamics which is hardly possible in the framework of classical 
continuum mechanics of viscous fluids. We then discuss a possible application of the theory to 
modeling of turbulent flows. Herewith, we propose a viewpoint according to which the turbulence is 
treated as a \textit{non-equilibrium state} emerging from the laminar (equilibrium) state as a 
result of non-linear interplay between 
\textit{dissipation} and \textit{dispersion}.

Despite the basic model~\cite{HPR2016,DPRZ2016,DPRZ2017,HYP2016} is applicable 
to both fluid and solid dynamics problems, its use 
in the solid dynamics context has been known for half a century already since the works of Godunov 
and Romenski in 1970s~\cite{GodRom1974,God1978,Romenski1979}, see also historical remarks 
in~\cite{Hyper-Hypo2018}. Therefore, in this paper, we mainly 
focus on the fluid dynamics applications and, in particular, on turbulence modeling. However, to 
the 
best of our knowledge, the proposed system of governing equations is the first genuinely 
\textit{non-linear dynamical} model which can be used to describe dynamics of continuously 
distributed defects in solids 
(dislocations, disclinations). So far, either linear dynamical or non-linear static theories have 
been developed starting from the works of Nye~\cite{Nye1953}, Kondo, Bilby~\cite{Bilby1955} and 
Kr\"oner~\cite{Kroner1963}, e.g. see a historical review 
in~\cite{Yavari2012} (see also \cite{Kosevich1965,KadicEdelen1983,Grinyaev2000}).

Apart of the conventional mass density $ \rho $, momentum density $ \mm = \rho \vv$ ($ \vv $ is 
the material velocity), entropy 
density $ s $ 
and total energy density $ E $, another important field of our theory is the so-called distortion 
field $ \Dist = [\triad{a}{i}] $,
$ a,i=1,2,3 $. The distortion field can be viewed as the 
\textit{local basis triad} attached to each 
material element. Thus, the evolution of $ \bm{A} $ provides complete information about the 
deformation and rotation of material elements. For further discussion, it is important to 
emphasize 
that the material elements are assumed to have finite length scale $ \ell $ and 
thus, are deformable, that is they posses a certain structure in contrast to the classical point of
view according to which the material elements are treated as scale-less points. For example, for 
Newtonian fluids, the length scale $ \ell $ can be 
estimated as $ \ell \sim \tau \csh $~\cite{HYP2016}, where $ \tau 
$ is the characteristic dissipation time scale (characteristic time for the material element 
rearrangements, e.g. see~\cite{HYP2016,DPRZ2016}), and $ \csh $ is the characteristic velocity of 
propagation 
of \textit{shear} perturbations.

\begin{figure}
	\begin{center}
		\begin{tabular}{ccc}
		\textbf{(a)} 
		\includegraphics[draft=false,width=0.30\textwidth]{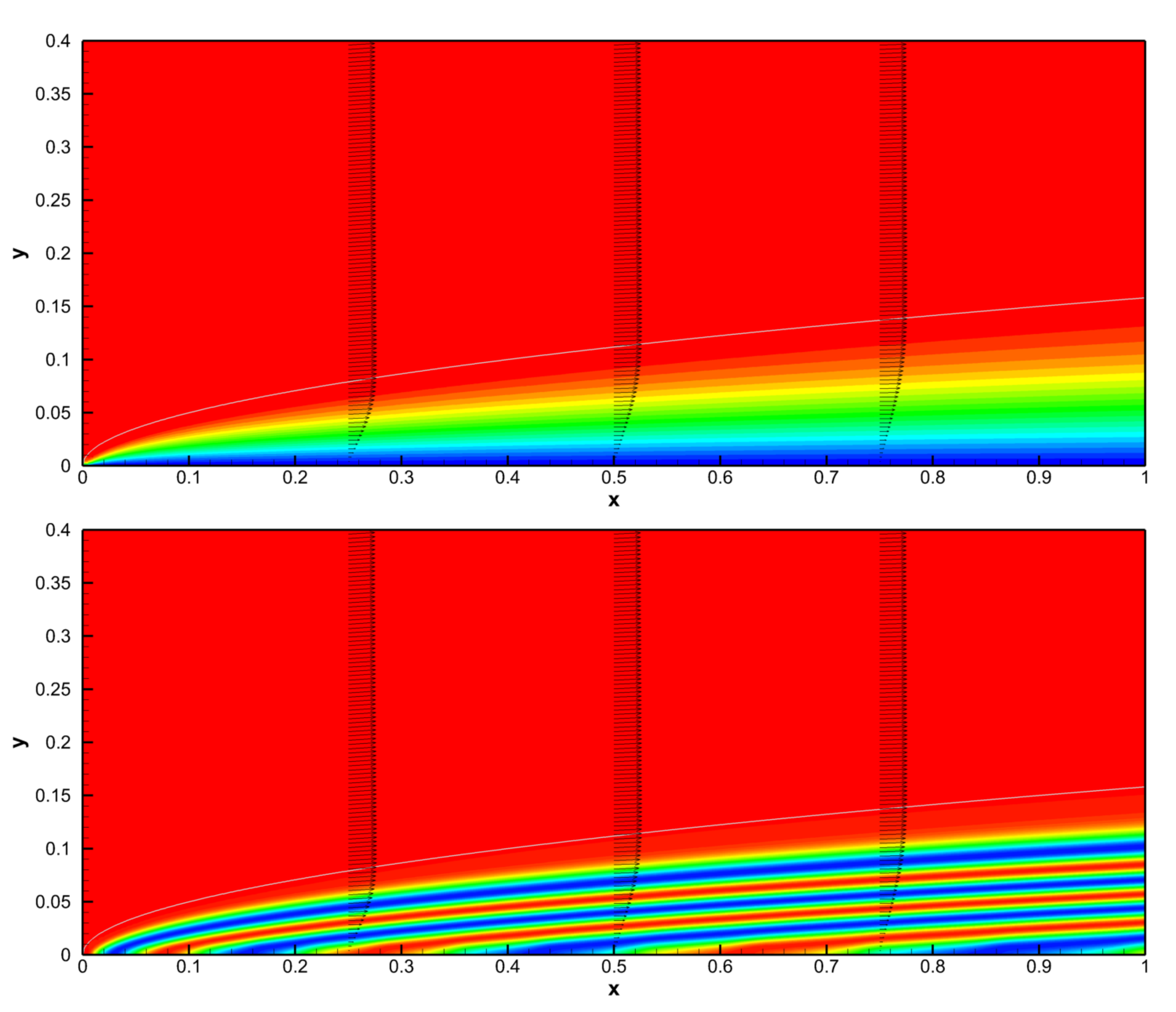} &
		\textbf{(b}) 
		\includegraphics[draft=false,width=0.25\textwidth]{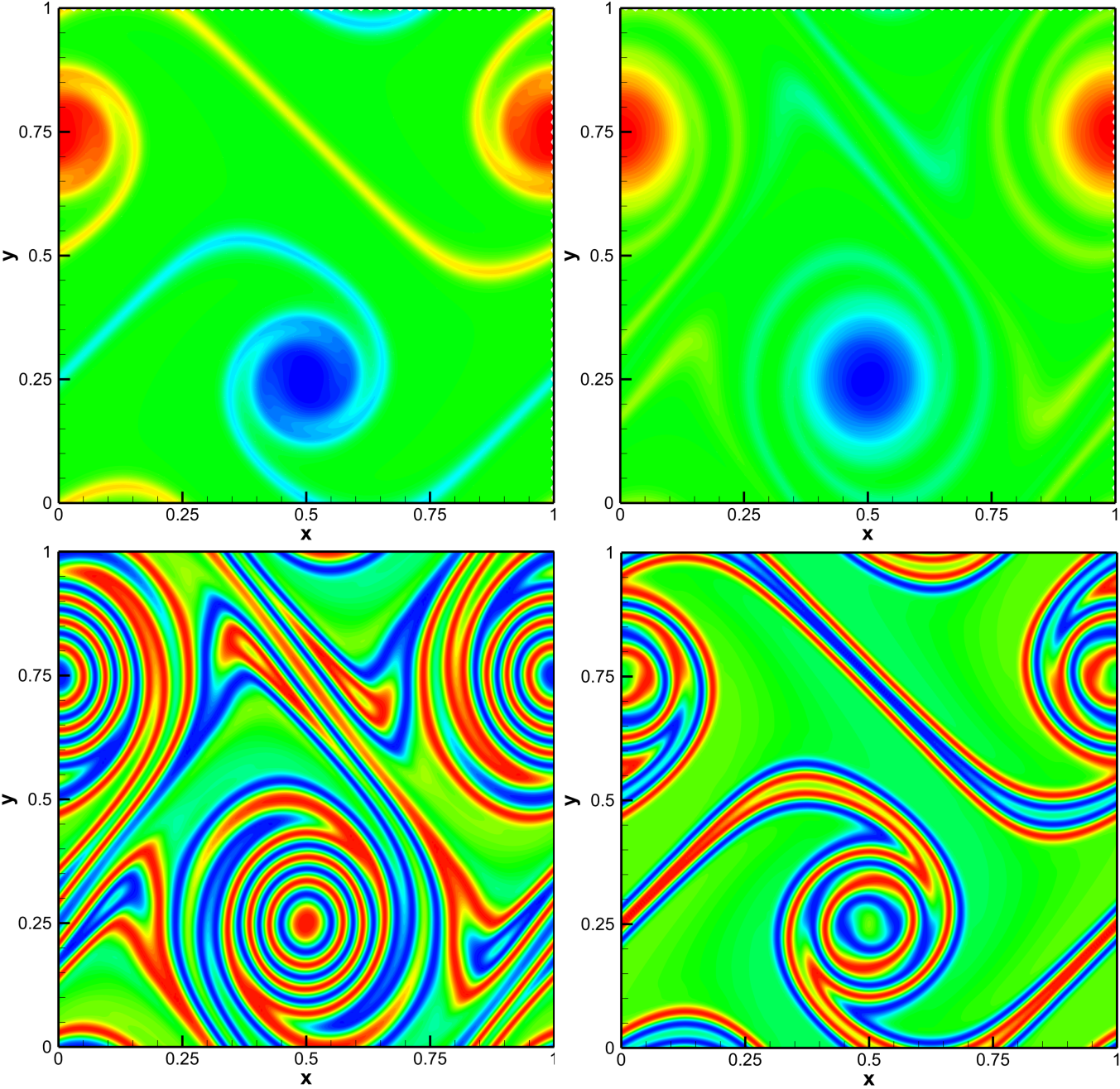} &
		\textbf{(c)}
		\includegraphics[draft=false,width=0.25\textwidth]{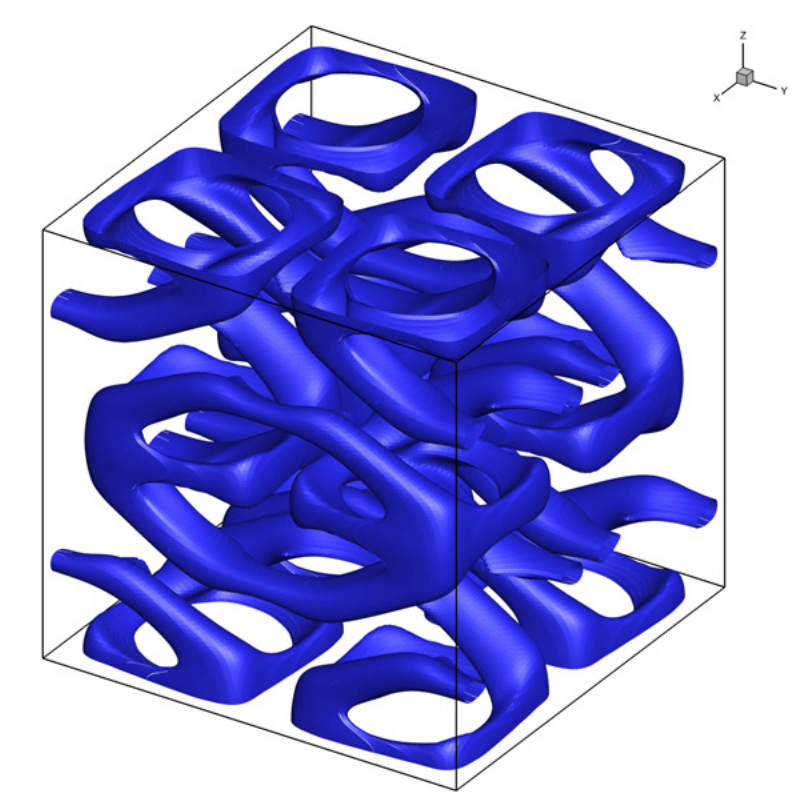} 
		\end{tabular} 
	\end{center}
	\vspace{-8mm}
	\caption{{\footnotesize \it (a): Laminar boundary layer for a Newtonian fluid
	(inflow is from the left, the wall is at the bottom), see details in~\cite{DPRZ2016}. The 
	horizontal 
	velocity profile 	(top) 
	identical to the Navier-Stokes and is homogeneous. However, the distribution of the
	{internal rotations} (spin) is heterogeneous (bottom, $ A_{11} $ component). (b): double 
	shear layer problem at two time instants
		(columns), from~\cite{DPRZ2016}.
		Vorticity (top raw). Apparently, the distortion field (bottom, $ A_{11} $ component)
		provides more information about the flow structure than available in the velocity field
		(vorticity). (c): 3D Taylor-Green vortex, $ A_{11} $ component, see details 
		in~\cite{DPRZ2016}. }}
	\label{fig1}
\end{figure}

Treatment of the material elements as the finite volumes assumes that the material elements 
may have rotational degrees of freedom in general. This means that the material elements may 
rotate 
independently enough relative to each other. It is important to understand that such a spin is not 
related to the velocity field via vorticity because the distortion field $ \Dist $ and the fluid 
velocity $ \vv $ are 
two 
independent state variables. Thus, the internal spin of the fluid elements was clearly 
observed in the extensive numerical studies~\cite{DPRZ2016,DPRZ2017} of our unified theory in the 
context of 
Newtonian flows. For example, Fig.\ref{fig1},~a) shows the numerical solution of the laminar 
boundary layer problem for a viscous gas. The top figure depicts the horizontal velocity 
distribution which is \textit{homogeneous} and perfectly matches the Navier-Stokes 
solution~\cite{DPRZ2016}. On the 
other hand, the bottom figure in Fig.\ref{fig1}~a) shows $ A_{11} $ component of the distortion 
field which is clearly \textit{heterogeneous} and is due to the rotation of the triad $ \AA $. 
Such 
a heterogeneous distribution cannot be derived by any means from the homogeneous distribution of 
the velocity 
field of the Navier-Stokes solution. Fig.\ref{fig1}~b) compares the vorticity distribution (top) 
with one of the components of the distortion field (bottom). Much more flow structure is visible 
in 
the distortion field. Fig.\ref{fig1}~c) shows a 3D structure in the three-dimensional Taylor-Green 
vortex problem ($ A_{11} $ component is shown), see details in~\cite{DPRZ2016}. We note that so 
far 
the spin of the triad field was 
computed as a byproduct in our simulations and it does not affect the solution due to the gauge 
freedom in the relation between the distortion field and the stress tensor, see 
Section\,\ref{sec.basic}. It 
is thus a principal goal of this paper to account for the material element spin in a 
thermodynamically consistent way.

Note that in the presence of the material spin, the basis triad field $ \bm{A} $ becomes 
\textit{anholonomic triad} (see definitions in Section~\ref{sec.diffgeom}), which means that it has
the non-vanishing torsion tensor $ \pd_i \triad{a}{k} - \pd_k \triad{a}{i}  $, where $ \pd_i $
are
partial space derivatives. It also means that the \textit{internal geometry} of the flow is
non-Euclidean, see Section~\ref{sec.diffgeom}.
Recall that the use of anholonomic basis triads (or actually tetrads) is well known in such 
geometrical 
theories as 
Einstein-Cartan theory of relativity~\cite{Friedrich1976} or teleparallel 
equivalent of general relativity 
(TEGR)~\cite{KleinertMultivalued,AldrovandiPereiraBook,Cai2016,Golovnev2017a} in which 
the
gravity
interaction is due to not only the curvature of the spacetime but also due to torsion which is 
coupled to the intrinsic angular momentum (spin) of the matter. This paper, therefore, can be 
considered as an attempt to look at the turbulence problem from the standpoint of classical 
geometrical field theories such as general relativity or Yang-Mills gauge 
theory~\cite{BaezMuniain}.

Finally, note that, in principle, one may want to try to account for the rotational \dof of the 
distortion 
$ \Dist $ via the polar decomposition $ \Dist = \bm{R}\sqrt{\GG} $, where $ \bm{R} $ is the 
rotation matrix, and $ \GG = \Dist^\transpose\Dist $ is the deformation tensor. Such an approach 
was developed by the Cosserat brothers~\cite{Cosserat1909,Forest2013}. In the 
non-linear settings as in this paper, 
the time evolution for $ \bm{R} $ can be obtained \cite{Chen1993a,Rosati1999} and $ \bm{R} $ 
can be considered as a natural choice of the state variable to represent the internal rotations. 
However, in the context of multiphysics modeling and development of efficient computational 
methods this choice is not that obvious. For example, the PDE for $ \bm{R} $ is a nonlinear 
equation with no apparent structure. It is therefore not clear if it can be derived from a 
variational principle or if it admits a Hamiltonian formulation via Poisson brackets 
\cite{SHTC-GENERIC-CMAT}. Without such a 
formulation, it is not clear how the PDE for $ \bm{R} $ can be consistently (mechanically and 
thermodynamically) 
coupled with other 
physical fields, for example, in the presence of electromagnetic forces, non-equilibrium mass and 
heat transfer, etc. 
On the other hand, a variational or Hamiltonian formulation for a model provides a quite universal 
way for model extensions towards nonlinear coupling with other physical fields. 
Therefore, instead of following the original Cosserat approach to describe internal rotational DoF 
via the polar decomposition of $ \Dist $, 
we follow a less straightforward (at first sight) but, in fact, more geometrical 
and elegant
way proposed by Cartan~\cite{Cartan1986,Hehl2007,Delphenich2016} who, in fact, was inspired by the 
work of 
Cosserat 
brothers 
and 
developed a continuum model based on Riemann-Cartan manifold (see historical remarks 
in~\cite{Scholz2018}), endowed 
with not only the 
metric (measuring the shape change) but also a non-symmetric affine connection with non-vanishing 
torsion (a measure of anholonomy). It has appeared that the use of Cartan's 
geometrical approach 
allows us to 
build a non-linear 
thermodynamically and mechanically consistent model with internal spin which shares some common 
features with the Yang-Mills gauge theory, posses a variational formulation, and has a rather 
elegant structure of governing equations with good  mathematical properties (symmetric 
hyperbolicity, well-posedness of the initial value problem). Also, most likely these governing 
equations also have a Hamiltonian structure (i.e they can be generated by the corresponding Poisson 
brackets) but this requires some further considerations and will be done elsewhere. Also, we 
note that the presence of a clear structure of governing equations is beneficial for the 
development of the structure-preserving numerical methods (i.e. when the PDE structure is respected 
at the discrete level), e.g. symplectic integrators \cite{Morrison2017}.

\section{Basic SHTC equations for fluids and solids}\label{sec.basic}

The basic SHTC system describing flow of a continuous medium governs the 
evolution of the mass 
density $ \rho $, momentum
density $
\bm{m} = \rho \bm{v} = [m_i] $ (where $ \vv = [\vel{i}]$ is the medium velocity), entropy density 
$ 
s 
$, the 
distortion field $ \bm{A} $, and the total 
energy density $ E $  and can be 
written in a Cartesian coordinate system as
\begin{subequations}\label{PDE.basic}
\begin{align}
	& \frac{\pd \rho}{\pd t} +        \pd_k(\rho \vel{k}) = 0,
	\\[2mm]
	& \frac{\pd m_i}{\pd t} +         \pd_k \left( m_i \vel{k} + p \kronecker{k}{i} + \dist{a}{i} 
	E_{\dist{a}{k}} \right) = 0,\label{PDE.basic.m}
	\\[1mm]
	& \frac{\pd \dist{a}{k}}{\pd t} + \pd_k (\dist{a}{i} \vel{i}) + \vel{j} \left(\pd_j \dist{a}{k} 
	- 
	\pd_k\dist{a}{j}\right)  = - 
	\theta^{-1} E_{\dist{a}{k}},\label{PDE.basic.A}
	\\[1mm]
	& \frac{\pd s}{\pd t} +           \pd_k (s \vel{k}) =
	\frac{1}{E_s \theta} E_{\dist{a}{i}} E_{\dist{a}{i}} \geq 0,
	\\[1mm]
	& \frac{\pd E}{\pd t} +           \pd_k \left( E \vel{k}  + \vel{i}  (p \, \kronecker{k}{i} + 
	\dist{a}{i} E_{\dist{a}{k}}) \right) = 0,
\end{align}
\end{subequations}
where the notations  $ E_\rho $, $ E_s $, $
E_{\dist{a}{i}} $, etc.
stand for $ \frac{\pd E}{\pd \rho} $, $ \frac{\pd E}{\pd s} $, $ \frac{\pd E}{\pd \dist{a}{i}} $
and have
the
meaning of thermodynamic forces, 
$ p = \rho E_\rho + s E_s +m_i E_{m_i} -
E = \rho^2
\varepsilon_\rho$
is the thermodynamic pressure, $ E = E(\rho,s,\vv,\AA) $ is the total energy density and should be 
provided in order to close the system, $ E_s = T $
is the temperature,  $
\varepsilon(\rho,s) $ is the internal specific energy, $ \theta \sim \rho \tau 
\csh^2 $
where $ \csh$ is the mentioned above shear sound speed, and $ \tau $ is the 
relaxation time scale 
which, in our approach, replaces the classical viscosity coefficient
and characterizes the ability of the medium to
flow. For simple flows the energy can be
taken as 
\begin{equation}\label{energy.basic}
E = \rho \, \varepsilon(\rho,s) + \rho \frac{\csh^2}{4} ||\bm{G}'||^2 + \frac{1}{2\rho}||\mm||^2,
\end{equation}
where $ \bm{G} =\AA^\transpose\AA$, and $ \bm{G}' = \bm{G} -
\frac{\text{tr}(\bm{G})}{3}\bm{I} $ is the deviatoric part of $ \GG $.
%, and $ ||\mm|| = \sqrt{\sum_{i}m_i } $ and $ ||\bm{G}'|| = \sqrt{\sum_{i,j}G'_{ij}} $.

The key parameter of the system \eqref{PDE.basic} controlling the transition from the solid to 
fluid regime is the relaxation time $ \tau $.
Thus, the medium responses elastically if the imposed perturbations have characteristic time scale 
$ T^{macro} $ much shorter than the relaxation time, $ \tau  \gg T^{macro} 
$~\cite{DPRZ2016,Boscheri2016,Hyper-Hypo2018}. The medium flows like a 
Newtonian fluid if 
$ \tau \ll T^{macro} $, and the medium response is of a visco-elastic type if $ \tau $ and $ 
T^{macro} $ 
are of comparable orders. One may think that the same is true for the Navier-Stokes-based fluid 
dynamics where in some simulations the solid state is represented by an infinite (very large) 
viscosity. This however leads to an infinitely rigid \textit{undeformable} solid. In contrast, the 
solid 
state in our approach is deformable and can be as soft as needed. The softness is controlled by 
the 
shear sound speed $ \csh $ (or shear modulus $ \rho \csh^2 $). 

Therefore, by providing a suitable model for $ \tau(\rho,s,\AA) $, one can obtain the whole range 
of fluid and solid regimes in one simulation. This is, in particular, an attractive feature of the 
model for 
problems where the solid and fluid state coexist. For example, granular 
flows~\cite{Pouliquen2006,Forterre2013}, flows of viscoplastic fluids~\cite{FrigaardReview2014}, 
melting and solidification problems in metallurgy and additive manufacturing (3D printing) 
\cite{Megahed2016,Mukherjee2017a,Yan2018}, polymer processing, etc.

An extensive validation of the model in the Newtonian regime and comparison against the 
Navier-Stokes equations has been done in \cite{DPRZ2016}. It was demonstrated analytically and 
numerically that Navier-Stokes solutions are indeed realized in system~\eqref{PDE.basic}. In 
particular, it was shown that for the choice $ \theta = \frac{1}{3 |\AA|^{5/3}}\rho \tau 
\csh^2 $, $ |\AA| =\det(\AA)$ of relaxation source term (the source of anholonomy) in the 
distortion 
PDE~\eqref{PDE.basic.A} 
and energy $ E $ of the form~\eqref{energy.basic},  
the effective shear viscosity is
\begin{equation}\label{viscosity}
\mu = \frac{1}{6}\rho \, \tau \csh^2 .
\end{equation}
Hence, the 
Reynolds number 
\begin{equation}\label{def.Re}
\Re = \frac{\rho v L}{\mu} \sim \frac{L}{\ell}
\end{equation}
scales as the ratio of the macroscopic length scale  $ L $ to the dissipation length scale $ \ell 
\sim \tau \csh $.

\section{Macroscopic observer}\label{sec.observer}

\begin{figure}
	\begin{center}
		\includegraphics[draft=false,width=0.35\textwidth]{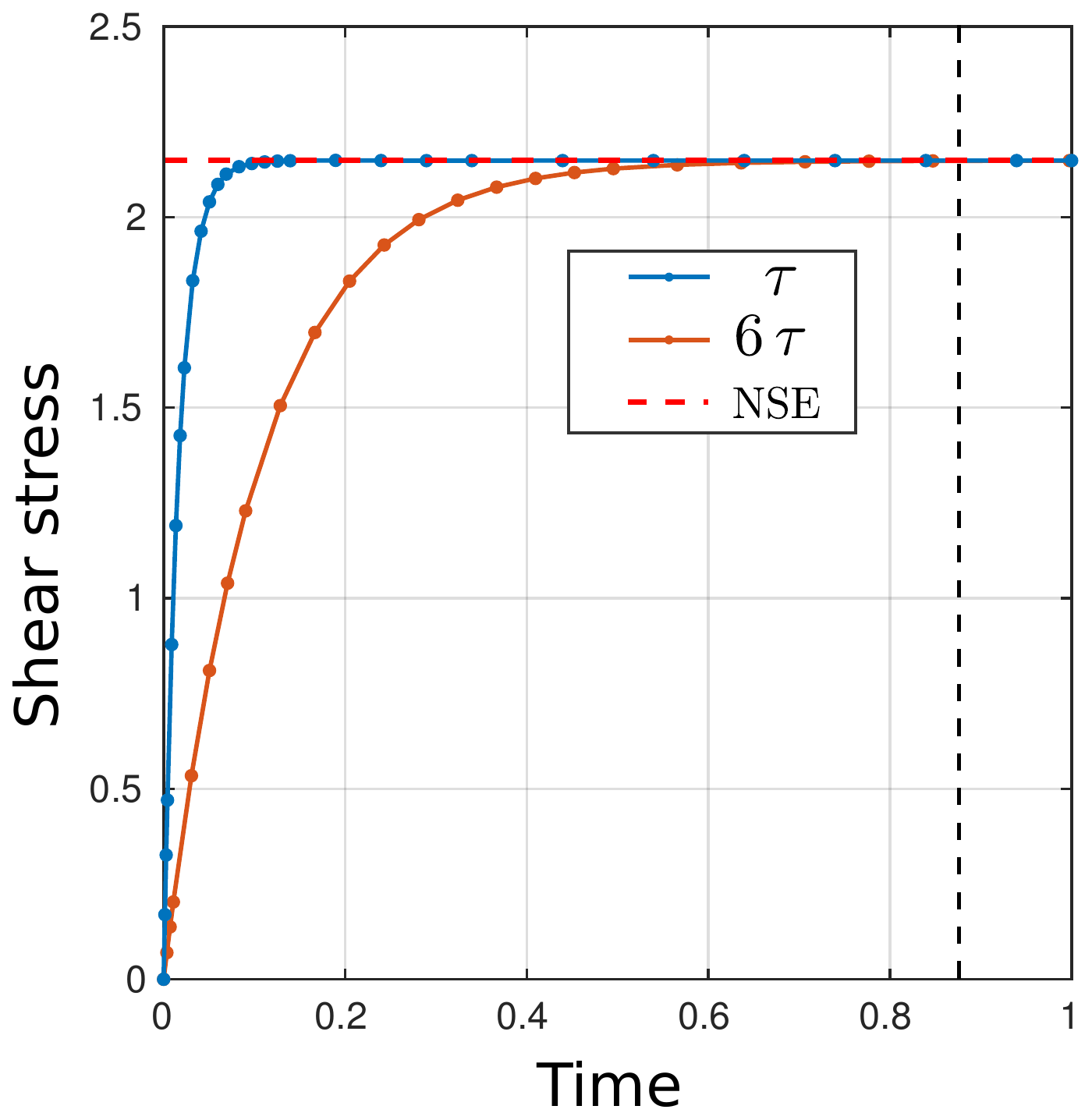}
	\end{center}
	\vspace{-8mm}
	\caption{{\footnotesize \it Time integration of the shear stress in a simple shear problem for 
	two values of the relaxation time $ \tau $ (details of the integration can be found 
	in~\cite{HPR2016}). The horizontal dashed line (red) is the 
	Navier-Stokes solution. The vertical dashed line (black) depicts the characteristic time scale 
	of the 
	macroscopic observer. Two curves are indistinguishable for the observer since they both provide 
	the same 
	stress level while the differences in the stress state are present at the smaller time scales.}}
	\label{fig2}
\end{figure}

Despite the physical values for $ \tau $ and $ \csh $ can be recovered for a given fluid from sound 
wave propagation data~\cite{HYP2016,brazhkin2012two,bolmatov2013thermodynamic,
bolmatov2015revealing,Bolmatov2015a,Bolmatov2016} they might be irrelevant for the practical use 
of 
the model~\eqref{PDE.basic} for Newtonian flows of weakly viscous fluids such as air or 
water\footnote{Note that the 
difference between gases and liquids is encoded in the hydrodynamic part $ \veps(\rho,s) $ of the 
total energy left unspecified in~\eqref{energy.basic}. For instance, one may use the ideal gas 
equation of state for gases, while the stiffened gas or Mie-Gr\"uneisen equation of state give a 
good approximation for liquids.}. For example, typical values of $ \tau $ may range from $ 10^{-7} 
$~s to $ 10^{-10} $~s for fluids with shear viscosity of the order $ 10^{-3} - 10^{-5} $~Pa$ \cdot 
$s which, if applied to problems having $ T^{macro} $ even of the order of seconds, makes the 
system \eqref{PDE.basic} extremely stiff. The good news is that the solutions corresponding to 
different relaxation times $ 
\tau' = 6\mu/(\rho \csh'^2) $ and $\tau'' = 6\mu/(\rho \csh''^2) $ (say $ \tau' = 
10^{-10} $~s and $ \tau'' = 10^{-3} $~s) but to the 
\textit{same value} of the effective shear viscosity $ \mu = \rho \tau' \csh'^2 /6= \rho \tau'' 
\csh''^2/6 $ are 
\textit{practically indistinguishable} as long as both relaxation times satisfy $ \tau' \ll 
T^{macro} $ and $ \tau'' \ll T^{macro} $. In this case, the macroscopic observer sees the same 
stress state, 
e.g. see the stress relaxation profiles in Fig.~\ref{fig2}. Obviously, the corresponding 
dissipation 
length scales $ 
\ell' = \tau' \csh' $ and $ \ell'' = \tau'' \csh'' $ are also different ($ \csh' $ and $ \csh'' $ 
should be computed from \eqref{viscosity}). Therefore, one may rightfully question what is then 
the 
physical meaning of the length scale $ \ell $ if it may differ by several orders of magnitude?

In order to address this issue, we reject using the physical values for $ \tau $ in practical 
simulations of
Newtonian flows and follow the methodology similar to those adopted in Large Eddy Simulations 
(LES) 
of 
turbulent flows and in 
particular in implicit LES (ILES)~\cite{Margolin2014}. Thus, following Margolin's 
terminology~\cite{Margolin2014}, we introduce the notion of the \textit{macroscopic observer} 
which 
is ``a length scale $ L $ that separates what is known or can be measured about a flow from what 
is 
unknown or unresolved''. Importantly, that the observer length is not determined by the flow, but 
by the questions that are being asked about the flow and so varies from problem to problem.
In practice, this is the computational cell size $ L = \Delta x $. The events that happen at 
scales 
smaller than $ \Delta x $ (or equivalently, faster than the corresponding time scale) are not 
available for a \textit{direct observation} by the macroscopic 
observer but emerge as indirect evidences in macroscopic quantities (e.g. momentum density) and 
have to 
be modeled. 
%For example, the quantities that can be observed are the standard mass, momentum and 
%energy densities. 

We then define the relaxation time $ \tau $ such that, for \textit{laminar flows}, 
the corresponding dissipation 
length scale 
$ \ell < L = \Delta x$ is the largest length scale for 
which Newtons' law of viscosity still provides a good approximation of the stress state inside the 
computational cells of characteristic 
length $ \Delta x $. In the terminology of~\cite{Margolin2014}, $ \ell $ 
is the 
\textit{physical length scale} at which dissipation is significant. Note that such an adaptive 
strategy for selecting the relaxation time scale is similar to the one adopted by Nishikawa in his 
``hyperbolic Navier-Stokes solver''~\cite{Nishikawa2011a,Nishikawa2017}.

Once the hierarchy of basic length scales has been defined, we may focus on the main goal of the 
paper, that is describing how the spin of the material elements can be taken into account in a 
thermodynamically compatible way. But first, we provide some motivation from the differential 
geometry in the following section.

%For the further understanding, it is important to emphasize that $ \AA $ and velocity $
%\vv $ are
%\textbf{independent} fields, e.g. $ \AA $ cannot be recovered from the velocity gradient
%or
%vorticity. This
%can be clearly seen
%in Fig.\ref{fig2} and \ref{fig3}. Thus, the rotational
%degrees of freedom of $ \AA $ is of the key interest in this paper. So far, these degrees of
%freedom was only passively taken into account in our simulations, i.e. it was computed 
%automatically as a byproduct because the distortion field is within the state variables. The idea 
%we want to investigate in this and the following papers is whether such rotational degrees of 
%freedom in the distortion field can be used to account for the kinetic energy of small-scale 
%eddies 
%in the turbulence modeling or dislocation dynamics in solids deformed plastically.

\section{Kinematics of the continuum from the differential geometry standpoint}\label{sec.diffgeom}

This section intends to recall basic definitions from the Riemannian and non-Riemannian 
differential geometry and provides a 
motivation for how the distortion spin can be taken into account. Also, from what follows, it 
will be 
clear that the distortion field can be treated as the field of Cartan's moving frame and thus, 
provides a basis for introduction of the internal 
non-Riemannian (Riemann-Cartan) flow geometry.

We treat the flowing continuous medium (either fluid or solid) as a 3D material manifold $ \Matter 
$ 
evolving in 
time $ t $
in the three-dimensional Euclidean space $ R^3 $. For simplicity, the ambient Euclidean space is 
equipped with the 
Cartesian coordinate system $ x^i $, $ i=1,2,3 $, that is the metric $ g_{ij} $ of $ R^3 $ is 
constant and identified with the Kronecker delta $ 
\delta_{ij} $ (in principle, arbitrary curvilinear coordinate systems $ x^i $ can be considered 
with a location dependent metric $ g_{ij}(x^k) $). 
In what follows, we shall consistently use letters $ i,j,k = 1,2,3 $ to index objects living in 
the 
manifold $ \Matter $ and its \textit{stressed} tangent space $ \TM $, capital letters $ 
\sA,\sB,\sC 
= 1,2,3 $ to 
index objects living in the 
\textit{globally} relaxed (globally stress-free) manifold $ \MatterO $ and its tangent space $ 
\TMgr $ at a point $ \myxi{A} $, 
and $ a,b,c = 1,2,3 
$ to index objects in the \textit{locally} relaxed tangent space (defined below) $ \TMlr $ at a 
point 
$ x^i $ of $ \Matter $.

\subsection{Holonomic case}

There are two intrinsic notions in the Lagrangian viewpoint on the continuum mechanics. 
Firstly, it 
is implied that it is possible to introduce a field of labels $ \myxi{A} $, $ \sA = 1,2,3 $ which 
label individual material elements and do not change in time. The material labels $ \myxi{A} $  
form the coordinate system in the globally relaxed matter manifold denoted as $ \MatterO $ and are 
called  the
Lagrangian coordinates. Usually, they can be identified with the location of 
the 
medium with respect to the coordinate system $ x^i $ at the initial time instant $ t=0 $. 
Secondly, the laws of motion, or simply the motion, of the continuum is identified 
with a map
\begin{equation}\label{def.motion}
x^i = \motion{i}(t,\myxi{A}),
\end{equation}
which, for every time instant $ t $, establishes a one-to-one correspondence between the manifolds 
$ \MatterO $ and $ \Matter $. The motion $ 
\motion{i}(t,\myxi{A}) $ thus induces naturally a 
coordinate system on the manifold $ \Matter $. Such a coordinate system is deforming with the 
medium and 	is curvilinear in general. 
Furthermore, it induces the coordinate, or \textit{holonomic}, basis $ \bm{\pd}_i $ 
and the dual basis (coordinate co-basis) $ \bm{\rmd \chi}^i $ which constitute the bases in the  
\textit{stressed} tangent 
space $ \TM $.
Similarly, at every point $ \myxi{A} $ of $ \MatterO $, there is a tangent space, called 
\textit{globally relaxed} 
tangent space $ \TMgr $ with the coordinate basis $ \bm{\pd}_{\sA} $ and co-basis $ 
\bm{\rmd \xi}^{\sA} $ induced by the Lagrangian coordinate system $ \myxi{A} $.

Because the motion $ \motion{i} $ is a single-valued map (i.e. its inverse $ 
\myxi{A}(t,\motion{i}) 
$ exists), its gradients $ \defgrad{i}{A} := \frac{\pd \motion{i}}{\pd\myxi{A}}$ and $ 
\idefgrad{A}{i} := \frac{\pd\myxi{A}}{\pd\motion{i}}$ are well-defined at every time instant. 
Thus, 
the bases of the tangent spaces $ \TM $ and $ \TMgr $ are related by
\begin{subequations}\label{holonom.triads}
\begin{equation}
\defgrad{i}{A} \bm{\rmd \xi}^{\sA} 	= \bm{\rmd \chi}^i,\qquad
\bm{\rmd \xi}^{\sA} 				= \idefgrad{A}{i} \bm{\rmd \chi}^i,
\end{equation}
\begin{equation}
 \bm{\pd }_{\sA} 				= \defgrad{i}{A}\bm{\pd}_i, \qquad
 \idefgrad{A}{i}\bm{\pd }_{\sA} = \bm{\pd}_i.
\end{equation}
\end{subequations}

Furthermore, in order to measure distances on $ \Matter $ and $ \MatterO $, one needs a metric. 
For 
isotropic media, it is naturally to assume that the globally relaxed manifold $ \MatterO $ is 
initiated with 
the metric $ g_{\sA\sB} $ which coincides with the metric $ g_{ij} $ of the ambient space (in 
general, the 
specification of the material metric $ g_{\sA\sB} $ should be done based on the physical 
properties 
of the medium, e.g. anisotropy, presence of residual stresses, etc.~\cite{Sedov1965}) and hence,
the \textit{material metric} $ G_{ij}(t,x^k) $ on $ \Matter $ is 
defined as
\begin{equation}\label{def.matter.metric}
G_{ij} := \frac{\pd\myxi{A}}{\pd\motion{i}}\frac{\pd\myxi{B}}{\pd\motion{j}} g_{\sA\sB} = 
\idefgrad{A}{i}\idefgrad{B}{j} g_{\sA\sB},
\end{equation}
which in general (if the motion is not a pure rotation or translation) does not equal to the space 
metric $ g_{ij} $. The material metric $ G_{ij} $ thus defines the distances on $ \Matter $ and 
can 
be used to construct other strain measures, e.g. $ E_{ij} := \frac{1}{2}(g_{ij} - 
G_{ij}) $.

For the further discussion, it is important to remark that, because $ \idefgrad{A}{i} $ is the 
gradient of $ \myxi{A}(t,\motion{i}) $, the partial 
derivatives $ 
\pd_i = 
\frac{\pd}{\pd \motion{i}} = \frac{\pd}{\pd x^i}$ commute\footnote{The partial derivative notation 
$ \pd_i $ should not be confused with the coordinate bases vector notation $ \bm{\pd}_i $ (written 
in bold) 
of the curvilinear coordinates $ \motion{i} $. Remark that the coordinate bases $ \bm{\pd}_{x^i} $ 
and $ \bm{\pd}_{\motion{i}} = \bm{\pd}_i $ of the coordinates $ x^i $ and $ \motion{i} $, 
correspondingly, are different in general.}
\begin{equation}
\pd_i \idefgrad{A}{j} - \pd_j \idefgrad{A}{i} = \pd_i\pd_j \myxi{A}(x^k) - \pd_j\pd_i 
\myxi{A}(x^k) 
\equiv 0,
\end{equation}
that is the quantity, called \textit{torsion tensor} below (see \eqref{def.torsion}),
\begin{equation}
T^i_{\ jk} := \defgrad{i}{A}(\pd_j \idefgrad{A}{k} - \pd_k \idefgrad{A}{j}) \equiv 0
\end{equation}
identically vanishes. In this case, the gradients $ \defgrad{i}{A} $ and $ 
\idefgrad{A}{i} $ are called \textit{holonomic 
triads}, while the coordinate bases $ \bm{\pd}_i $, $ \bm{\pd}_{\sA} $ and co-bases $ \bm{\rmd 
\chi}^i $, $ \bm{\rmd \xi}^{\sA} $ are called holonomic frames and co-frames, respectively.

The above description intrinsically relies on the notion of the motion $ x^i = 
\motion{i}(t,\myxi{A}) $ as a 
single-valued map between the two manifolds $ \MatterO $ and $ \Matter $. Such a description, 
however, is 
limited and can be only applied to describe reversible deformations of the continuum. In order to 
be able to deal with irreversible deformations (flows), that is to deal with the \textit{material 
element rearrangements}, the one-to-one correspondence between $ \MatterO $ and $ \Matter $ should 
be ``destroyed''. The material element rearrangements implies that the single-valued map $ 
\motion{i}(t,\myxi{A}) $ should be replaced with a multi-valued and/or non-smooth 
map~\cite{KleinertMultivalued,Rakotomanana}. In fact, it is rather natural to assume that the 
current state of a flowing medium should not depend on the entire flow history encoded in the laws 
of motion~\eqref{def.motion}. This implies that the map $ x^i = \motion{i}(t,\myxi{A}) $ and the 
globally relaxed material manifold $ \MatterO $ should be 
eliminated from the mathematical formulation of the governing equations for media deforming 
irreversibly. As we shall demonstrate in what follows, such a theory requires the 
replacement of the holonomic triads~\eqref{holonom.triads} by \textit{anholonomic 
triads} with non-zero torsion which constitute the basis of the Riemann-Cartan geometry.
%This has a simple motivation, that is the response of a flowing 
%medium may not depend on the entire motion history.

\subsection{Non-holonomic case}

Geometrically speaking, the motion of the continuum described in the previous section can be 
viewed 
as the tangent bundle with two tangent spaces at each point of the base space, which is $ 
\Matter 
$. The fiber $ \TM $ is a natural tangent space spanned by the holonomic triad $ 
\bm{\pd}_i $. The second  
tangent space is $ \TMgr $ spanned by the triad $ \bm{\pd}_{\sA} $. It can be viewed not as the 
tangent space to $ \MatterO $ at a point $ \myxi{A} $ but directly as a tangent space to the 
base space $ \Matter $ \textit{soldered} at the point $ x^i = \motion{i}(t,\myxi{A}) $. 

Such a construct of a manifold with two tangent spaces is conditioned by the following reasoning. 
In principle, it is possible to introduce an evolving in time locally relaxed manifold, say $ 
\MatterR $, instead of 
the globally relaxed manifold $ \MatterO $, even in the case of a medium deformed irreversibly. 
However, due to the material element rearrangements, it is no longer possible to establish a 
global one-to-one correspondence between the manifolds $ \Matter $ and $ \MatterR $. The absence 
of 
such a map implies that the geometry of $ \MatterR $ is \textit{incompatible} with the Euclidean 
geometry of the ambient space and $ \MatterR $ cannot be realized in $ R^3 $ (otherwise the 
geometry of $ \MatterR $ would be equivalent to the Euclidean geometry). This also can be  
illustrated by the following thought experiment. 
If one partitions the material manifold $ \Matter $ into small pieces (material elements) along 
the 
coordinate lines $ x^i $ and allows each piece individually and \textit{instantaneously} to relax, 
it would be impossible to realize the relaxed material as a whole from the relaxed pieces 
(the pieces would not fit to each other globally without gaps). 
%Because we do not assume that there is a one-to-one correspondence between the manifolds $ 
%\Matter 
%$ and $ \MatterR $, that is there is no real motion transforming one manifold into another, the 
%triad $ \triad{a}{i} $ is not a gradient of a map, like $ \idefgrad{A}{i} = \frac{\pd 
%\myxi{A}}{\pd 
%\motion{i}} $ is the gradient of $ 
%\myxi{A}(t,\motion{i}) $.
%Thus, the relaxed 
%manifold $ \MatterR $ is incompatible with the geometry of Euclidean space, that is there are
%\textit{incompatibilities}. 
However, one may imagine that the small relaxed material elements lie tightly in a non-Riemannian 
manifold with nonzero curvature and/or torsion (and even non-metricity)~\cite{Yavari2012}.
Despite the global geometry of the locally relaxed manifold $ \MatterR $ is not accessible (e.g. 
we 
do not need and actually, we do not know how
to introduce a coordinate system on it), its local geometrical structure can be still recovered 
via 
the 
availability of the tangent spaces to $ \MatterR $ which, however, are 
now soldered to the manifold $ \Matter $ directly, and a proper definition of affine connection. 
It 
has appeared that this is pretty enough to build a consistent theory of irreversible deformation.
We emphasize that the detailed knowledge of the geometry of $ \MatterR $ is not of interest on its 
own, but it is necessary to define the stress field in the flowing medium.

Motivated by this reasoning, we shall consider the flowing medium as the manifold $ \Matter $ with 
two tangent spaces at each point $ x^i $ with the only difference that the holonomic frames $ 
\bm{\pd}_i $ and $ \bm{\pd}_{\sA} $ will be replaced by the anholonomic frames $ \ee_i $ and $ 
\ee_{a} $. The tangent space spanned by the anholonomic basis $ \bm{e}_i $ is called the 
\textit{stressed tangent space} and denoted as $ \TM $. The tangent space spanned by the 
anholonomic basis $ \bm{e}_{a} $ (recall that the objects living in this space are indexed with 
small Latin letters $ a,b,c =1,2,3 $) is called the \textit{locally relaxed tangent space} and 
denoted as $ \TMlr $.

We then introduce the field of 
\textit{non-coordinate} (\textit{or 
anholonomic}) \textit{basis triad} $ \triad{a}{i} $ and its inverse $ \itriad{i}{a} $ as the 
matrices of a
linear 
map between the anholonomic co-frame fields $ \ee^{i} $ and $ \ee^{a} $ 
%(it is of course implied that $  
%\det(\triad{a}{i}) > 0 $ and hence, the inverse triad $ \itriad{i}{a} $ exists such that $ 
%\itriad{i}{a}\ee_i = \ee_a $)
\begin{equation}\label{def.cobasis}
\itriad{i}{a}\ee^a = \ee^i, \qquad \ee^a = \triad{a}{i}\ee^i.
\end{equation}
Here, the co-frame field $ \ee^i $ is dual to $ \ee_i $ and $ \ee^{a} $ is dual to $ \ee_{a} $. 
In the case of isotropic media, there is a metric field $ g_{ab}(x^i) $ defined on $ \Matter $ 
which is 
used to 
measure lengths in the relaxed tangent space $ \TMlr $. The metric $ g_{ab} $ is not a dynamical 
parameter of the theory and is assumed to be 
globally constant, $ g_{ab}(x^i) = \delta_{ab} 
$ so that $ \ee_a\cdot\ee_b = g_{ab} $ and
$ \ee^a\cdot\ee^b = g^{ab} $ with $ g^{ab} $ being the inverse of $ g_{ab} $. 
The distortion field defined as the solution to PDE \eqref{PDE.basic.A} is identified with the 
anholonomic triads $ \triad{a}{i} $ in \eqref{def.cobasis}. In the following, we shall directly 
work with the matrices $ \triad{a}{i} $ and $ \itriad{i}{a} $ instead of the co-frames $ \ee^i $ 
and $ \ee^a $.

At each point of $ \Matter $ there is also another metric field $ G_{ij} $ which is used to 
measure 
lengths in the stressed tangent space $ \TM $ and which is defined analogously to 
\eqref{def.matter.metric}
\begin{equation}\label{material.metric}
G_{ij} = \triad{a}{i}\triad{b}{j} g_{ab}.
\end{equation}
It is the metric $ G_{ij} $ and not $ g_{ab} $ which should be used to measure 
distances on $ \Matter $.
Note that such a metric is defined up to an arbitrary local rotation $ \Rot{a}{a}(x^i) $ of the 
triad $ A^{a'}_{\ i} = \Rot{a}{a} (x^i) \triad{a}{i} $ in the tangent space $ \TMlr $, 
i.e. $ 
G_{ij} = 
g_{ab}\triad{a}{i}\triad{b}{j} = 
g_{ab} \Qot{a}{a}A^{a'}_{\ i} \Qot{b}{b}A^{b'}_{\ j} = g_{a'b'}A^{a'}_{\ i}A^{b'}_{\ j}$, where $ 
\Qot{a}{a}$ is the inverse of $ \Rot{a}{a} $. Thus, if one is interested in only the deformation 
DoF of the 
material elements then such rotational DoF of the triad field should 
be considered as redundant (gauge invariance). However, it is the main objective of this paper to 
try 
to 
relate such a gauge invariance in the dynamics of the triad field (distortion field) to the 
rotational 
DoF of the microstructure of the flowing medium. Thus, in 
what 
follows, we shall discuss how the presence of rotational DoF of the triad field defines the 
non-Riemannian geometry of the relaxed material manifold $ \MatterR $ which, in turn, defines the 
stress field in the medium.

%In particular, it is not implied that, in general, the co-frame field $ \ee^a $ can be triad can 
%not be associated to a gradient of 
%
%
%as a frame of basis vectors obtained from the coordinate basis $ \bm{\pd}_i $ by a 
%linear map with the matrix $ \itriad{i}{a} $, $ 
%\det(\itriad{a}{i}) > 0 $. Since $ \itriad{i}{a} $ is assumed to be a 
%non-degenerate matrix, the inverse $ \triad{a}{i} $ can be defined, $ 
%\itriad{i}{a}\triad{a}{j} = 
%\delta^i_{\ j}$ and $ \itriad{i}{a}\triad{b}{i} = \delta^a_{\ b}$, so that the vectors of dual 
%bases (co-bases) $ 
%\bm{e}^a $ and $ \bm{\rmd x}^i $ are connected as
%\begin{equation}\label{def.cobasis}
%\itriad{i}{a} \bm{e}^a = \bm{\rmd x}^i, \qquad \bm{e}^a = \triad{a}{i} \bm{\rmd x}^i.
%\end{equation}

%We now proceed in defining the internal geometry of the flowing continuum. 
%We shall assume that the relaxed manifold $ \MatterR $ is flat, i.e. it is equipped with the 
%Euclidean metric $ g_{ab} = \delta_{ab} $, i.e the triads $ \bm{e}^a $ and $ \bm{e}_a $ are 
%orthonormal, $ 
%\ee_a\cdot\ee_b = g_{ab}$. 

We now proceed in defining the geometry of $ \MatterR $. However, because $ \MatterR $ is not 
explicitly available, we shall work directly on the manifold $ \Matter $ and use the fact that the 
tangent spaces to $ \MatterR $ are now soldered to $ \Matter $. This means that the objects 
defined 
in the tangent spaces $ \TMlr $ and $ \TM $ with respect to the frames $ \ee_a $ and $ \ee_i $, 
respectively, can be locally transformed from one space into another in the manner of 
\eqref{material.metric} using the triads $ \triad{a}{i} $ and $ \itriad{i}{a} $, e.g. for the 
vector 
fields $ v^a(x^i) $ and $ v_a(x^i) $ defined in $ \TMlr $ we have their representations $ 
v^i(x^i) $ and $ \vel{i}(x^i) $ in $ \TM $
\begin{equation}
v^i = \itriad{i}{a} v^a, \qquad \vel{i} = \triad{a}{i} v_a.
\end{equation}

So far, we have defined only metric $ g_{ab}(x^i) $ which only partially characterizes the 
geometry of $ \MatterR $. Note that, despite $ g_{ab}(x^i) = const $ globally, it is, in general, 
not enough to conclude that $ \MatterR $ is flat\footnote{If one, however, works in the settings 
of 
the Riemannian geometry, $ g_{ab}(x^i) = \delta_{ab} $ is, of course, equivalent to the flatness 
of 
the manifold.} in the settings of non-Riemannian geometry. In addition, one needs to define an 
affine 
connection in order to define the parallel transport and thus, the covariant differentiation of 
the 
vector fields on $ \MatterR $.

Because the orthonormal bases $ \ee^a $ are the only admissible bases in the tangent space $ \TMlr 
$, the local rotations $ \Rot{a}{a}(x^i) $ form the symmetry group, or the gauge group, on $ \TMlr 
$ of 
admissible 
transformations 
of the bases. Hence, $ \Omega^a_{\ bc} $ defined as 
\begin{equation}\label{spin.connection}
\Omega^a_{\ bc} := R^a_{\ a'}\pd_b Q^{a'}_{\ c}
\end{equation}
is a natural connection in $ \MatterR $ which is also called the \textit{spin connection} and 
which 
defines the covariant differentiation of a contravariant $ v^a(x^i) $ and covariant 
vector field $ v_a(x^i) $ as
\begin{equation}
\nabla_a v^b = \pd_a v^b + \Omega^b_{\ ac} v^c, \qquad \nabla_a v_b = \pd_a v_b - \Omega^c_{\ ab} 
v_c,
\end{equation}
where $ \pd_a $ is nothing else but
\begin{equation}\label{deriv.transform}
\pd_a := \itriad{i}{a}\pd_i.
\end{equation}

In this paper, however, we shall consider only a particular realization of such a geometry. 
Namely, 
we shall assume that the frames $ \ee^a(x^i) $ are all fixed to be globally aligned which implies 
that the only admissible transformations of the frame field are global (point-independent) 
rotations $  
\Rot{a}{a}(x^i) = const $. The rotation of the frame field $ \ee^i $ is thus measured with respect 
to 
the chosen global orientation of the frame field $ \ee^a $. Fixing a preferable global frame 
results in 
that the spin connection $ \Omega^a_{\ bc} $ vanishes in this frame
and the covariant derivatives $ \nabla_a $ coincide with $ \pd_a $, $ \nabla_a = \pd_a $. The 
general case with no preferable orientations of the frames and thus with non-vanishing spin 
connection will be considered in a later publication. In this case, the local rotation filed $ 
\Rot{a}{a}(x^i) $ should play the role of an independent state variable with its own time 
evolution 
equation. In principle, this introduces an extra degree of freedom and add extra inertia. 

Since we are intending to develop an Eulerian system, i.e. written in the coordinates $ x^i $, it 
is necessary to rewrite the covariant derivatives $ \pd_a v^b $ and $ \pd_a v_b $ (written in the 
co-basis $ \ee^a $) in the co-basis $ \ee^i $. Thus, one obtains
\begin{equation}\label{cov.deriv.transform}
\pd_a v^b = \itriad{i}{a}\triad{b}{j} \nabla_i v^j, \qquad \pd_a v_b = \itriad{i}{a}\itriad{j}{b} 
\nabla_i \vel{j},
\end{equation}
where the covariant derivatives $ \nabla_i $ are defined as
\begin{equation}\label{def.cov.diff}
\nabla_i v^j = \pd_i \, v^j + \W{j}{ik}v^k, \qquad \qquad \nabla_i \vel{j} = \pd_i \, \vel{j} - 
\W{k}{ij}\vel{k},
\end{equation}
with $ \W{i}{jk} $ being the so-called \Weitz connection
\begin{equation}\label{def.Weitzenbock}
\W{i}{jk} := \itriad{i}{a}\pd_j \triad{a}{k}.
\end{equation}

An important property of the \Weitz connection is that if the 
triad field $ \triad{a}{i} $ is anholonomic then $ \W{i}{jk} $ is not symmetric in the lower 
indices 
and 
hence, it has the non-vanishing \textit{torsion tensor} (the field strength of the triad) which is 
defined as 
\begin{equation}\label{def.torsion}
\tors{i}{jk} := \W{i}{jk} - \W{i}{kj} = \itriad{i}{a} (\pd_j\triad{a}{k} - \pd_k\triad{a}{j})
\end{equation}
and can be viewed as the measure of anholonomy. If rewritten in the co-basis $ 
\ee^a $, the torsion $ \tors{i}{jk} $ is
\begin{equation}\label{def.torsion.ab}
\tors{i}{ab} = \pd_a \itriad{i}{b} - \pd_b \itriad{i}{a} = -\itriad{j}{a}\itriad{k}{b} 
\tors{i}{jk} 
=-\itriad{j}{a}\itriad{k}{b}\itriad{i}{c} 
(\pd_j\triad{c}{k} - \pd_k\triad{a}{j}).
\end{equation}
Therefore, even though the metric $ g_{ab}(x^i) $ is globally constant, $ \MatterR $ has
non-zero torsion $ T^i_{\ ab} $ in general. This means that the geometry of the manifold $ 
\MatterR 
$ is non-Euclidean. The manifold with non-zero torsion (and curvature) constitute the subject of 
the Riemann-Cartan geometry. In such a theory, the manifold is said to be flat, or Euclidean, if 
only 
both the torsion 
and curvature are zero. Therefore, the manifold $ \TMlr $ having the constant metric $ g_{ab}(x^i) 
= 
const $ but non-zero torsion $ T^i_{\ ab} $ is not flat.

Another important feature of the \Weitz connection is that the curvature tensor (the field 
strength 
of the connection)
\begin{equation}\label{def.cruv.tens}
R^i_{\ jkl} = \partial_k \W{i}{lj} - \partial_l \W{i}{kj} + \W{i}{km}\W{m}{lj} -  
\W{i}{lm}\W{m}{kj}
= \itriad{i}{a}(\pd_j\pd_k \triad{a}{l} - \pd_k\pd_j \triad{a}{l})  \equiv 0
\end{equation}
identically vanishes, e.g. see~\cite{KleinertMultivalued,FeckoBook,Yavari2012}. This, however, 
happens due to fixing of a global preferable orientation of the frames $ \ee^a(x^i) $ in $ \TMlr 
$. In the general case,  when the spin connection $ \Omega^a_{\ bc} $ does not vanish, the 
curvature tensor does not vanish too. This case will be considered elsewhere.

Finally, note that the triads $ \triad{a}{i} $ and $ \itriad{i}{a} $ are \textit{covariantly 
constant}, 
i.e. they are the fields for which the covariant derivative in an arbitrary direction vanishes
\begin{equation}
\nabla_j \triad{a}{i} = \pd_j \, \triad{a}{i} - \W{k}{ji}\triad{a}{k} = 0, \qquad 
\nabla_j \itriad{i}{a} = \pd_j \, \itriad{i}{a} + \W{i}{jk} \itriad{k}{a} = 0.
\end{equation}
Not that, if written in the frame $ \ee^a $, the triad $ \triad{a}{i} $ and $ \itriad{i}{a} $ 
become $ \delta^a_{\ b} $ and hence, the vanishing of the covariant derivatives of triads can also 
be seen from the transformation rule~\eqref{cov.deriv.transform} and the fact that $ \pd_a 
\delta^b_{\ c} = 0 $.
This means that the manifold $ \MatterR $, despite being non-flat, possess \textit{absolute 
parallelism}, 
or 
\textit{teleparallelism} (i.e. parallelism ``at a distance''), allowing the path independent 
parallel transport of vectors~\cite{FeckoBook}. Also, note that due to \eqref{material.metric} and 
the assumption 
that $ g_{ab}(x^i) =\delta_{ab}$, the \Weitz connection has vanishing non-metricity $ \nabla_{i} 
G_{jk} = 
0 $, that is it is the \textit{metric compatible connection}. 

Remarkably, the discussed properties of the 
\Weitz 
connection is in the basis of the so-called teleparallell reformulations of Einstein's general 
relativity (the latter is a torsion free theory), e.g. 
see~\cite{KleinertMultivalued,AldrovandiPereiraBook,Cai2016,Golovnev2017a,Rakotomanana}. Thus,
the theory considered in this paper, i.e. with fixing the orientation for the frame field $ 
\ee^a $, is equivalent to the ``pure tetrad''  teleparallel 
gravity (in the terminology of~\cite{Golovnev2017a,Krssak2018}) while the general case with 
non-vanishing spin connection~\eqref{spin.connection} corresponds 
to 
the fully covariant version of teleparallel gravity. The 
geometrical 
settings of our theory may thus help one to see some analogies between the gravity interaction and 
interactions between distant points in a turbulent flow due to rotational DoF of unresolved 
scales.

We thus have discussed that the non-Riemannian geometry of the relaxed manifold $ \MatterR $ can 
be 
determined by defining a field of 
anholonomic frames on it (Cartan's moving frames). The evolution of non-Euclidean properties 
of 
such a geometry is 
then 
defined by the evolution of the spin intensity (torsion) of the basis triads. The 
motivation for this paper is thus to adopt the apparatus of the Riemann-Cartan geometry for 
describing (modeling) the internal  structure of turbulent flows whose nature is essentially in 
the 
rotational dynamics of small scales eddies.

In the rest of the paper, instead of $ T^i_{\ jk} $, we shall use its mixed-indices counterpart $ 
T^a_{\ jk} := \pd_j\triad{a}{k} - \pd_k\triad{a}{j} = \triad{a}{i} T^i_{\ jk}$, also called the 
\textit{field 
strength} of the triad~\cite{Cai2016}. Moreover, we shall restrict ourselves to 
the Cartesian coordinates $ x^i $ so that the placement of the indices (up or down) will not play 
an 
essential role. Thus, in 3D, there are only 9 independent components of the torsion $ T^a_{\ jk} $ 
\begin{equation}
\left (\begin{array}{ccc}
	     0       & T^a_{\ 12}  & T^a_{\ 13} \\
	-T^a_{\ 12} &      0       & T^a_{\ 23} \\
	-T^a_{\ 13} & -T^a_{\ 23} &      0
\end{array}\right),
\end{equation}
which can be cast to the 3-by-3 matrix
\begin{equation}
  \burg{a}{i} := \LeviCivitaUp{ijk} \pd_j \dist{a}{k},
\end{equation}
where $ \LeviCivitaUp{mjk} $ is the Levi-Civita symbol.

%
%The coefficient of anholonomy (don't know if we need it at all), it is defined in the tangent 
%space
%\begin{equation}
%\O{a}{bc} := \itriad{j}{b} \itriad{k}{c} (\pd_j\triad{a}{k} - \pd_k\triad{a}{j})
%\end{equation}

\section{Extended SHTC equations for 
media with internal spin}\label{sec.extended}

We now return to the modeling part and derive an extended system of PDEs which is 
the 
main result of this paper. Thus, in order to account for the rotational degrees of freedom of the 
distortion field $ \AA $, we shall use the torsion tensor, which, in 3D, reduces to a 3-by-3 matrix
\begin{equation}\label{torsion}
 \Burg = \alpha\nabla\times\AA \qquad\text{or}  \qquad \burg{a}{i} = \alpha \, 
 \LeviCivitaUp{ijk}\pd_j 
 \dist{a}{k}, 
\end{equation}
where $ \alpha \sim \length^{-1} $ is
just a
scaling constant, '$ \length $' is a length unit so that $ \Burg \sim 
\length^{-2} 
$. In 
the 
fluid dynamics 
context, the torsion field
can be
understood as the \textit{number density} of unresolved vortex lines piercing through unit areas 
oriented normal to the coordinate directions, while in the solid dynamics context it is treated as 
the number  density of continuously 
distributed defects 
such as dislocations and disclinations\footnote{Note that in the dynamics of solids, linear 
theories 
of 
defects are usually used, e.g. see historical review in~\cite{Yavari2012}. In the linear regime, 
it 
is possible 
to separate rotational degrees of 
freedom of the distortion field from the pure deformation part (displacement) and thus, it is 
possible to use two tensors one of which represents the density of dislocations (curl of the 
displacement field) while the second 
one represents the density of disclinations (curl of the rotation), e.g. 
see~\cite{VolovichKatanaev1992}. In a non-linear 
formulation like the one used in this paper, such a separation is hardly possible in general 
because 
both degrees of freedom, the rotation and deformation, are coupled multiplicatively which can be  
seen in the polar decomposition of the 
distortion, see 
Section\,\ref{sec.intro}.}~\cite{GodRom2003,VolovichKatanaev1992,Guzev2000,KleinertMultivalued,Yavari2012}.
Before to discuss the evolution equation for $ \Burg $ let us first recall our 
motivation for the choice of $ \Burg $ to represent internal rotations instead of the 
conventional micropolar rotation $ \bm{R} = \Dist \GG^{-1/2}$ and its gradients $ 
\nabla\bm{R} $ 
(micropolar curvature), e.g. 
see 
\cite{Steinmann1994,Ehlers2018}. Indeed, the matrices $ \bm{R} $ and $ \nabla\bm{R} $ contain only 
the information 
about the rotations of the triad field while $ \Burg $ also incorporates some excessive  
information (contained in $ \GG $) on the stretch of the triads. At first sight, this may say in 
favor of utilizing $ \bm{R} $ as the proper micropolar state variable. However, our long-term goal 
is to build a consistent nonlinear multi-physics continuous theory for modeling of internal 
rotations in the presence of electromagnetic fields or non-equilibrium mass and heat transfer, and 
it has appeared that the torsion field $ \Burg $ is more beneficial in the multi-physics modeling 
context. This is due to the fact that governing PDEs for $ \bm{R} $ and $ \nabla\bm{R} $ are very 
nonlinear and do not have an apparent structure of Euler-Lagrange equations and therefore, their 
variational nature is unknown. Without a variational formulation, a consistent extension of 
the conventional micropolar continuum towards nonlinear coupling with new fields is 
unclear. On the other hand, the governing equation for $ \Burg $, as it is discussed below, is 
simpler and admits a variational formulation. Therefore, our preferences are given to those state 
variables which have a better structure of the governing equations while their intuitive  
interpretation might be less evident. Another important argument in favor of $ \Burg $ is that it 
comes with the underlying geometrical structure of the Riemann-Cartan geometry as it has been 
discussed in the previous section.

The 
time 
evolution for $ \Burg
$ can be
obtained~\cite{SHTC-GENERIC-CMAT,GodRom2003} easily by means of applying the curl operator to 
\eqref{PDE.basic.A}. As the result, one gets
%\begin{equation}\label{PDE.Burg}
%\frac{\pd \Burg}{\pd t} + \nabla\times(\Burg\times\vv + \alpha\theta^{-1}E_{\AA}) +
%\vv\otimes(\nabla\cdot\Burg)
%= 0.\end{equation}
\begin{equation}\label{PDE.Burg}
\frac{\pd \burg{a}{i}}{\pd t} + \frac{\pd}{\pd x^k} \left(
\burg{a}{i} \vel{k} - \vel{i} \burg{a}{k} + \LeviCivitaUp{ikj} \alpha \theta^{-1} E_{\dist{a}{j}}
\right) + \vel{i} \frac{\pd \burg{a}{k}}{\pd x^k} = 0.
\end{equation}

It, in
particular,
follows
immediately from this PDE that the torsion is an intrinsic property
of the viscous dissipation in fluids or irreversible deformation in solids. Indeed, if the 
dissipative source term in \eqref{PDE.basic.A} is absent, $
\theta^{-1}E_{\AA} \equiv
0  $, then, from \eqref{PDE.Burg}, we automatically have $ \Burg \equiv 0 $ for all time
instants if it
was zero  initially.

%\begin{remark}
%In principle, the material element
%rotations can be also described via the introduction of the intrinsic angular
%momentum~\cite{Shliomis1967,Eringen1972,Berezin1995,Heinloo2004,Sadovskii2013}.
%This,
%however, results in the non-symmetric stress tensor. The use of the torsion allows
%to avoid the appearance of the non-symmetric stress tensor because the torsion represents 
%relative 
%rotations of the 
%material elements with
%respect to each other. However, the torsion tensor itself does not describe the rotation of $ 
%\Dist 
%$ directly but it does it an indirect way. Indeed, the torsion simply indicates if the
%neighboring triads $ \Dist $ are compatible ($ \Burg = 0$) or not ($ \Burg \neq 0 $). But the 
%time 
%rate of the torsion does describe the rate of change of the incompatibility in $ \Dist $ and 
%thus, 
%the 
%distortion spin. As we shall see, in this way, the resulting stress tensor in a medium with the 
%intrinsic spin remains symmetric.
%\end{remark}

%In principle, the distortion can be decomposed as $ \Dist = \Rot\sqrt{\Gsf}$ (polar 
%decomposition), 
%where $ \Rot $ is the 
%orthogonal matrix, i.e. $ \Rot^\transpose\Rot = \Id $, and represents the rotation of the triad $ 
%\Dist $, 
%while $ \Gsf = \Dist^\transpose\Dist $ describes the deformation of the material element. 
%However, 
%the PDE 
%for $ \Rot $ in the full nonlinear formulation is very complicated with no apparent structure. It 
%is therefore unlikely that it can be derived from the Hamilton principle!

\begin{remark} Note that even though $ \nabla\cdot \Burg = 0 $ (as follows automatically from 
the definition \eqref{torsion}), the term $ \vel{i} \pd_k \burg{a}{k} $
should be retained in \eqref{PDE.Burg} in order to preserve the proper characteristic
structure and Galilean/Lorentz invariance property~\cite{DPRZ2017,SHTC-GENERIC-CMAT}.
\end{remark}

So far, PDE \eqref{PDE.Burg} does not add any new physics to the basic model because it does not 
contribute neither to the momentum flux nor energy. To relate $ \Burg $ to a certain physics we 
have to 
treat it as a real
\textit{independent} state
variable, i.e. we should include $ \Burg $ into the set of arguments of the total energy. This, 
however, 
is not a straightforward task and should be done without violating the thermodynamic consistency 
of 
the basic model~\eqref{PDE.basic}. 

In this paper, we follow a route similar to those taken in the framework of SHTC 
equations~\cite{SHTC-GENERIC-CMAT}. Thus, we extend the set of state
variables, so that the new (extended) 
energy potential is 
\begin{equation}\label{energy.extend.gen}
\calE = \calE(\rho,s,M_i,\dist{a}{i},\burg{a}{i},\durg{a}{i}),
\end{equation}
where
$ M_i$ is the generalized momentum density, $ \burg{a}{i}$ is the torsion field, and $
\durg{a}{i}$ is a field \textit{complimentary} in a certain sense to $ \burg{a}{i} $. 
In 
fact, as we shall see in Section\,\ref{sec.Hamilton}, the fields $ \burg{a}{i} $ and $ \durg{a}{i} 
$ can be 
viewed as parts of the only one rank-3 four-torsion tensor. In Section\,\ref{sec.Hamilton}, one 
may, in particular, see some 
similarities 
between these fields and electric and magnetic fields in electrodynamics where the 
three-dimensional electromagnetic vector fields are also parts of the single four-dimensional 
rank-2 
electromagnetic tensor.

%Thus, the 
%extended 
%system we shall obtain has the 
%same structure as recently 
%obtained equations of electrodynamics in moving 
%media and can be 
%derived from the Hamilton 
%principle of stationary action\footnote{In the relativistic 
%literature, similar to the relativistic derivation of the Maxwell equations where the electric and 
%magnetic fields are just entries of only one electromagnetic tensor, the fields $ \Burg $ and $ 
%\Durg $ are 
%entries of the four-dimensional torsion $ B^a_{\ ij} $, e.g. see \cite{Maluf2013,Combi2017}.} in 
%which the distortion $ \Dist $ plays 
%the role similar to the 
%electrodynamic vector potential, $ \Durg $ plays the role equivalent to the electric field and 
%torsion plays the role similar to the magnetic field, see Section\,\ref{sec.Hamilton}. Thus, the 
%PDE for $ \Durg $ field can be 
%obtained as the 
%Euler-Lagrange equations, 
%while the torsion PDE \eqref{PDE.Burg} plays the role of the integrability condition for the PDE 
%for $ \Durg $.

Therefore, if one wishes to recover the SHTC structure of the extended system, a complimentary to $ 
\Burg $ 
field has to be introduced in such a way that the constitutive flux in \eqref{PDE.Burg}, $ \alpha \,
\theta^{-1} E_{\Dist}$, should be identified with the Legendre conjugate $ \calE_{\Durg} $ of the 
complimentary field $ \Durg $ so that \eqref{PDE.Burg} takes the form
\begin{equation}\label{PDE.Burg2}
\frac{\pd \burg{a}{i}}{\pd t} + \pd_k \left(
\burg{a}{i} \vel{k} - \vel{i} \burg{a}{k} + \LeviCivitaUp{ikj} \calE_{\durg{a}{j}}
\right) + \vel{i} \pd_k \burg{a}{k} = 0.
\end{equation}
After introducing fields $ \Burg $ and $ \Durg $, distortion 
PDE~\eqref{PDE.basic.A} reads
\begin{equation}\label{PDE.A.new}
\frac{\pd \dist{a}{k}}{\pd t} + \pd_k (  {\dist{a}{i} \vel{i}}  ) + 
\vel{j} \left(\pd_j \dist{a}{k} - 
\pd_k\dist{a}{j}\right) = -\alpha^{-1}\calE_{\durg{a}{k}}.
\end{equation}

\begin{remark}\label{rem.relax.limit}
It is important to remark that, in fact, we will not require the equality 
\begin{equation}\label{cond.equilib}
\alpha^{-1}\calE_{\Durg} = \theta^{-1}\calE_{\Dist}
\end{equation}
to be fulfilled at every time instants but we will construct the 
extended system in a more general way so that this equality is satisfied and thus, the basic 
model~\eqref{PDE.basic} is recovered, only in the relaxation limit of the extended model (i.e when 
a relaxation parameter goes to zero), see Section~\ref{sec.recovering.NS}.
\end{remark}

Furthermore, the time evolution for $ \Durg $ has to be provided. Actually, the PDE for $ \Durg $ 
is known. Thus, from the studying the 
structure of SHTC equations, the torsion PDE \eqref{PDE.Burg2} is consistent with the energy 
conservation if only it is accompanied by an equation for $ \Durg $ with a very precise 
structure, 
see~\cite{SHTC-GENERIC-CMAT}. Alternatively, this PDE can be derived 
from the variational principle, see Section\,\ref{sec.Hamilton}.

The system of governing equations thus reads
\begin{subequations}\label{PDE.extend}
\begin{align}
&\frac{\pd \rho}{\pd t} + \frac{\pd(\rho \vel{k})}{\pd x^k} = 0,\label{PDE.extend.rho}
\\[1mm]
&\frac{\pd M_i}{\pd t} + \frac{ \pd }{\pd x^k}  \left( M_i \vel{k} + P \kronecker{k}{i} + 
\dist{a}{i}
\calE_{\dist{a}{k}} - \burg{a}{k} \calE_{\burg{a}{ i}} - \durg{a}{k} \calE_{\durg{a}{ i}}
\right ) = 0,\label{PDE.extend.M}
\\[2mm]
&\frac{\pd \dist{a}{k}}{\pd t} +\frac{\pd (  {\dist{a}{i} \vel{i}}  )}{\pd x^k} + 
\vel{j} \left(\frac{\pd \dist{a}{k}}{\pd x^j} - \frac{\pd\dist{a}{j}}{\pd x^k}\right) = 
-\frac{1}{\alpha} \calE_{\durg{a}{k}},\label{PDE.extend.A}
\\[2mm]
&\frac{\pd \burg{a}{i}}{\pd t} + \frac{\pd}{\pd x^k} \left(
\burg{a}{i} \vel{k} - \vel{i} \burg{a}{k} + \LeviCivitaUp{ikj} \calE_{\durg{a}{j}}
\right) + \vel{i} \frac{\pd \burg{a}{k}}{\pd x^k} = 0,\label{PDE.extend.B}
\\[2mm]
&\frac{\pd \durg{a}{i}}{\pd t} + \frac{\pd} {\pd x^k} \left( \durg{a}{i} \vel{k}  -  \vel{i} 
\durg{a}{k} - \LeviCivitaUp{i k j}  \calE_{\burg{a}{j}} \right) + \vel{i} \frac{\pd 
\durg{a}{k}}{\pd x^k}  = \frac{1}{\alpha}\calE_{\dist{a}{i}} - 
\frac{1}{\eta}\calE_{\durg{a}{i}},\label{PDE.extend.D}
\\[2mm]
&\frac{\pd s}{\pd t} + \frac{\pd (s \vel{k})}{\pd x^k} =
\frac{1}{\calE_s \eta} E_{\durg{a}{i}} E_{\durg{a}{i}} \geq 0.\label{PDE.extend.s}
\end{align}
\end{subequations}
Here, the generalized momentum density $ \MM $
does not
equal to $ \mm = \rho\vv $ but may include the contribution from the torsion fields $ \Burg $ and  
$ \Durg $. 
For example in the
simplest case, which \textit{depends} on the energy specification (see 
equation \eqref{energy.torsion}
below), $
\MM =
\rho\vv
+ \Durg_a\times\Burg^a $ where $ \Durg_a $ and $ \Burg^a $ are $ a $-th
rows of $ \Durg $ and $ \Burg $ (summation over $ a $ is implied).
In general, the Cauchy stress tensor \begin{equation}
\Cauchystr{k}{i} = - P \kronecker{k}{i} - \dist{a}{i} 
\calE_{\dist{a}{k}} + \burg{a}{k} \calE_{\burg{a}{i}} + \durg{a}{k} \calE_{\durg{a}{i}}  
\end{equation} 
is non-symmetric ($ \Cauchystr{k}{i} \neq \Sigma_{i}^{\phantom{i}k} $) in media with internal 
rotations \cite{Steinmann1994}. Nevertheless, in order to guaranty the conservation of the total 
angular momentum (matter + fields, see the discussion in Section\,\ref{sec.ang.mom}) the full flux 
of the total 
momentum
\begin{equation}\label{momentum.flux}
\MomFlux{k}{i} = \vel{k} M_i -\Cauchystr{k}{i}
\end{equation}
has to be \textit{symmetric} because the equations \eqref{PDE.extend.M} represents not solely the 
balance 
of the 
macroscopic momentum $ \rho \vv $ but the conservation of the total momentum $ \MM $, see 
Section\,\ref{sec.ang.mom}. The symmetry 
of the momentum flux $ \MomFlux{k}{i} $ has 
to 
be taken into account in the construction of the energy potential $ \calE $, which is discussed in 
Section\,\ref{sec.closure}.
Also, $ P = \rho \calE_\rho + s\calE_s + M_i \calE_{M_i} + \burg{a}{i}\calE_{\burg{a}{i}} + 
\durg{a}{i}
\calE_{\durg{a}{i}} - \calE $ is the thermodynamic 
pressure which includes the contribution due to the torsion fields $ \Burg $ and 
$\Durg$.

\begin{remark}
As well as in the basic model \eqref{PDE.basic}, where we have $ \vv = E_{\mm} $, one can show 
that 
in the extended model \eqref{PDE.extend},
we have $ \vv = \calE_{\MM} $, that is the velocity is always the Legendre conjugate to the total 
momentum in the SHTC framework~\cite{SHTC-GENERIC-CMAT}.
\end{remark}

Note that the extra terms in the pressure and stress tensor in~\eqref{PDE.extend.M} are not 
arbitrary. 
Their appearance is conditioned by the 
structure of the evolution equations \eqref{PDE.extend.B} and \eqref{PDE.extend.D} and the 
requirement that the system should respect the energy conservation law. Thus, only this 
choice of the momentum flux is compatible with the energy conservation. Indeed, if one sums up the 
equations 
multiplied by the corresponding factors 
\begin{equation}\label{energy.sum}
\calE_\rho \cdot \eqref{PDE.extend.rho} + \calE_{M_i} \cdot \eqref{PDE.extend.M} + 
\calE_{\dist{a}{k}} 
\cdot \eqref{PDE.extend.A} + \calE_{\burg{a}{i}} \cdot \eqref{PDE.extend.B} + \calE_{\durg{a}{i}} 
\cdot 
\eqref{PDE.extend.D} + 
\calE_{s} \cdot \eqref{PDE.extend.s}
\end{equation}
the following energy conservation law can be obtained
\begin{equation}\label{energy.law}
\frac{\pd \calE}{\pd t} + \frac{\pd }{\pd x^k} \left( \vel{k} \calE + \vel{i} \Cauchystr{k}{i} + 
\LeviCivitaUp{ijk} \calE_{\durg{a}{i}}\calE_{\burg{a}{j}}\right) = 0,
\end{equation}
where the last term in the energy flux, $ \LeviCivitaUp{ijk} 
\calE_{\durg{a}{i}}\calE_{\burg{a}{j}} $, is the contribution due 
to torsion.

In the following section, we discuss the structure of the algebraic source terms on the right 
hand-side of~\eqref{PDE.extend}.

Lastly, we note that the left hand-side of the governing equations for $ \Burg $ and $ \Durg $ 
share a common structure with the equations of electrodynamics of moving 
media~\cite{DPRZ2017,SHTC-GENERIC-CMAT,Rom1998,Rom2001}. In particular, system \eqref{PDE.extend} 
can be symmetrized in the way discussed in \cite{SHTC-GENERIC-CMAT} and thus, its hyperbolicity 
depends on the convexity of the energy potential \eqref{energy.extend.gen}.
Nevertheless, we underline that the Hamiltonian structure (i.e. Poisson brackets generating the 
extended system) of system \eqref{PDE.extend} is not completely understood since it 
is not fully equivalent to the Hamiltonian structure of the electrodynamics equations due to the 
source 
terms $ -\alpha^{-1}\calE_{\Durg} $ and $ \alpha^{-1}\calE_{\Dist} $ which are reversible (i.e. 
they do not change sign with respect to the time reversal transformation 
\cite{Pavelka2014a,PKG-Book2018} in contrast to the irreversible term $ -\eta^{-1}\calE_{\Durg} $ 
in \eqref{PDE.extend.D}) and 
therefore, they are also a part of the reversible time evolution (i.e. they are a part of 
the corresponding Poisson brackets). This, however, requires further investigations and we plan to 
study the Hamiltonian structure of the extended model in detail in a subsequent publication.

\section{Irreversibility, dispersion and non-locality}

Recall that in the SHTC framework as well as in the more general GENERIC formulation of 
non-equilibrium thermodynamics~\cite{SHTC-GENERIC-CMAT,PKG-Book2018,Ottinger2005}, the 
reversible  
and 
irreversible\footnote{Here, we 
follow the definition of reversibility and irreversibility as given with respect to the 
\textit{time 
reversal transformation} (TRT), e.g. see \cite{Pavelka2014a,PKG-Book2018}. Thus, to distinguish 
between reversible and irreversible terms, we apply TRT to the evolution equations, and if a
term does not change its sign it is regarded to the reversible part.  The irreversible terms 
change 
their signs under TRT. } parts of the time 
evolution 
are treated separately. The reversible part conserves both the energy and entropy and can be 
derived from Hamilton's 
principle as well 
as can be generated from corresponding Poisson brackets~\cite{SHTC-GENERIC-CMAT}. The irreversible 
part of the time evolution raises the entropy but still conserves the total energy.

In particular, the reversible part of the time evolution of system \eqref{PDE.extend} 
is represented by the left hand-side and by the reversible source terms 
$ -\alpha^{-1}\calE_{\Durg} $ and $ \alpha^{-1}\calE_{\Dist} $. The irreversible part is 
represented by the source term $ \eta^{-1}\calE_{\Durg} $ in \eqref{PDE.extend.D} and $ 
(\calE_s\eta)^{-1}\calE_\Durg:\calE_\Durg $ in \eqref{PDE.extend.s}. Either reversible or 
irreversible \textit{source} terms 
are constructed in such a way that after the summation~\eqref{energy.sum}, their contribution to 
the energy conservation \eqref{energy.law} vanishes. 
%However, it has appeared that in order to 
%fulfill the 
%energy conservation, the use of only irreversible source terms is not enough for the extended 
%model 
%but also 
%\revOne{Q4.}{\textit{reversible}} \textit{source terms}, which \textit{conserve} the entropy, have 
%to be 
%introduced 
%in the system.   
Thus, the irreversible nature of the source term $ \eta^{-1}\calE_{\Durg} $ in \eqref{PDE.extend.D} 
is now more obvious because it is coupled with the 
source term in the entropy PDE~\eqref{PDE.extend.s} (entropy production). They annihilate 
after the  summation~\eqref{energy.sum}. On the other hand, the terms $ 
-\alpha^{-1}\calE_{\Durg} $ and $ \alpha^{-1}\calE_{\Dist} $ in \eqref{PDE.extend.A} and 
\eqref{PDE.extend.D} are indeed of the reversible nature and they annihilate with each other in the 
summation~\eqref{energy.sum} and thus, they 
do not contribute to the entropy production. 
%\sout{Such terms therefore, can be viewed as the dispersion source terms.
%The necessity to introduce these terms is conditioned by the 
%summation rule~\eqref{energy.sum}.}
The presence of the reversible source terms 
can 
be 
also justified from the physical ground. Thus, 
the extended system intends to account for impact of the small scale dynamics on the macroscopic 
dynamics and it  belongs to the class of \textit{non-local} models because the torsion is the 
first spatial
gradient of the distortion field while the field $ \Durg $ characterizes non-locality in time 
(micro-inertia), see 
Section\,\ref{sec.Hamilton}. In multi-scale systems with a strong inter-scale 
interaction, i.e. when the characteristic wave length of the perturbations is comparable with the 
characteristic length of the small scale flow structures, the dispersion phenomena are 
observed. Dispersive process can be viewed as process with \textit{reversible energy 
exchange} between the small and large scales. In the extended system~\eqref{PDE.extend}, the 
reversible source terms, therefore, responsible for the emergence of dispersion. For example, in 
the context of turbulence modeling, the 
reversible source terms can be 
related to the reversible energy exchange between large and small turbulent scales, the 
so-called \textit{energy cascade}. Remark that the antisymmetric-like structure of the reversible 
source 
terms is 
essentially the same as in the recently proposed first-order hyperbolic reformulation of the 
dispersive nonlinear 
Schr\"odinger equation by Dhaouadi et al~\cite{Dhaouadi2018}.

Note that in classical continuum mechanics, dispersive phenomena are usually modeled with high 
order PDE systems. However, recent results obtained with first-order hyperbolic 
equations~\cite{Romenski2011,Mazaheri2016,Dhaouadi2018} demonstrate that, in fact, dispersive 
phenomena can be also successfully modeled with first-order hyperbolic PDEs with relaxation type 
source terms.

\section{Closure: an example of equation of state}\label{sec.closure}

In order to close a system of PDEs in the SHTC framework, one has to provide the energy 
potential~\cite{SHTC-GENERIC-CMAT}. In particular, as it is clear from the structure of system 
\eqref{PDE.extend}, we have to define the total energy $\calE$ as the function of all the state 
variables in order to define constitutive fluxes and source terms. An additional 
constraint is that the provided energy potential should guaranty the 
symmetry of the total momentum flux $ \MomFlux{k}{i} $, see \eqref{momentum.flux}, in order to
guaranty the conservation of the total angular momentum, see Section\,\ref{sec.ang.mom}.

In this paper, we introduce only a particular but quite general example for $\calE$. 
Thus, we 
define $\calE$
\begin{equation}
\calE = \rho \veps(\rho,s) + \rho \frac{c_\text{sh}^2}{4} ||\GG'||^2 + \frac{1}{2\rho} \MM^2 +
\calE^t(\MM,\Burg,\Durg)
\end{equation}
as a sum of the basic energy \eqref{energy.basic} and an additional energy $ \calE^t $ which 
accounts for the contribution due to torsion, and can be identified with the energy contained in 
the small scale unresolved eddies of a turbulent flow or the energy carried by small scale defects 
in solids. As a starting point, one can start with 
\begin{equation}\label{energy.torsion}
%{\red \frac12 \rho\, c_t^2 ||\skewA||^2} + 
 \calE^t  = \frac{1}{2}\left (\frac1\epsilon \, 
 ||\Durg||^2 
 + 
 \frac1\mu \, ||\Burg||^2\right ) 
 -\frac{1}{\rho}\sum_{a=1}^{3}\left|
 \begin{array}{ccc}
 M_1 & \durg{a}{1} & \burg{a}{1} \\
 M_2 & \durg{a}{2} & \burg{a}{2} \\
 M_3 & \durg{a}{3} & \burg{a}{3}
 \end{array}
 \right|
  - \frac{\csp}{2}\sum_{a=1}^{3}\left|
  \begin{array}{ccc}
  \dist{a}{1} & \durg{a}{2} & \burg{a}{1} \\
  \dist{a}{2} & \durg{a}{2} & \burg{a}{2} \\
  \dist{a}{3} & \durg{a}{3} & \burg{a}{3}
  \end{array}
  \right|
\end{equation}
where the coefficients
$ \epsilon = \epsilon(\rho,s) 
$ and $ \mu = 
\mu(\rho,s) $ depend on $ \rho $ and $ s $, 
in general, but are assumed to be 
constant in this paper. They are some transport parameters 
such that they scale as square of velocity, $ (\epsilon\mu)^{-1}\sim v^2 $, see 
Section\,\ref{sec.units}. 
The parameter $ \csp $ has the dimension of velocity and characterizes the propagation 
of rotational degrees of freedom.
In general, 
one may try 
other energies $ \calE^t $, which, of course, will affect all terms in 
\eqref{PDE.extend} via the
thermodynamic forces $ \calE_{\Dist} $, $ \calE_{\Burg} $, $ \calE_{\Durg} $.

Note that if $ \calE^t $ is given by \eqref{energy.torsion}, the momentum density is ($ \Durg_a $ 
and $ \Burg^a $ are $ a $-th rows of $ \Durg $ and $ \Burg $, the summation over $ a $ is implied) 
\begin{equation}\label{momentum.total.example}
\MM =\rho \vv + \Durg_a\times\Burg^a
\end{equation}
and hence, the convective part $ v^k M_i $ of the momentum flux $ \MomFlux{k}{i} $, as 
well as the Cauchy stress tensor $ 
\Cauchystr{k}{i} $, is not symmetric. Nevertheless, the entire total momentum flux $ 
\MomFlux{k}{i}  = v^k M_i - \Cauchystr{k}{i}$ corresponding to the potential \eqref{energy.torsion} 
is \textit{symmetric} as can be verified by direct calculations.

%On the other hand, for the given 
%definition~\eqref{energy.torsion} of the energy $ \calE $, the 
%thermodynamic forces $ \calE_{\Burg} $ and $ \calE_{\Durg} $ and hence, the extra 
%stress, $ -\calE_{\Burg}\otimes\Burg - \calE_{\Durg}\otimes\Durg
%$, are also non-linearly coupled with the momentum density and thus, with the velocity of the 
%medium.

%\revOne{Q7}{$ \skewA = \frac12(\Dist - \Dist^\transpose) $ 
%is the skew-symmetric part of the distortion field, $ c_t $ is a microstructure characteristic 
%velocity. We note that if the energy $ \calE $ depends on $ \Dist $ only via the invariants of the 
%symemtric 
%tensor $ \GG = \Dist^\transpose\Dist $, the resulting stress $ \Dist^\transpose \calE_{\Dist} $ is 
%always symmetric. Therefore, the  term $ \frac12 \rho\, c_t^2 ||\skewA||^2 $ is directly 
%responsible for the loss of symmetry of the 
%overall stress tensor.} 

\section{Dimensional analysis}\label{sec.units}

In this section, we discuss the physical units of the parameters and state variables of the 
extended model. From this, we shall see that the new fields $ \Burg $ and $ \Durg $ have the 
meanings close to the number density and angular momentum (spin), respectively.

Let us denote the time unit as '$ \timeunit $', the length unit as '$ \length $', and the mass 
unit 
as '$\mass $'. It is convenient to chose the scaling parameter $ \alpha $ in \eqref{torsion} to 
scale as $ \length^{-1} $. In such a way, the torsion $ \Burg $ scales as $ \length^{-2} $ which 
can be interpreted as the number of dislocation lines (in solids) or vortex lines (in fluids) 
crossing a unit area perpendicular to each coordinate direction.

Thus, according to our choice $ [\alpha] = \length^{-1} $ and 
because $[\theta] = [\rho \tau \csh^2] = \frac{\mass}{\timeunit 
\cdot\length}$, we have
\begin{equation}
[\Burg] = \length^{-2}, \qquad [\calE_{\Durg}] = \length^{-1}\cdot\timeunit^{-1}.
\end{equation}
From that the quantity $ \calE_{\Durg}\otimes\Durg $ should have the dimension of stress, i.e. 
$\mass\cdot
\length^{-1}\cdot\timeunit^{-2} $, we get that
\begin{equation}\label{D.dim}
[\Durg] = \mass\cdot\timeunit^{-1}.
\end{equation} 
Thus, $ \Burg $ has the units of the number density while $ \Durg $ scales as the angular 
momentum, $ \mass\cdot\length^2\cdot\timeunit^{-1} $, divided by $ \length^{2} $. 
Also, it follows from \eqref{D.dim} that $ \Burg_i \times \Durg_i
\sim \rho\, \Vel $, i.e. it indeed scales as the momentum density, where $ \Vel = 
\length/\timeunit 
$ is 
some velocity.

Furthermore, transport coefficients $ \epsilon $ and $ \mu $ in \eqref{energy.torsion} and 
relaxation parameter $ \eta $ have the following units
\begin{equation}\label{dim.coef}
[\epsilon] = \mass\cdot\length, \qquad
[\mu] = \timeunit^2\cdot \length^{-3}\cdot\mass^{-1}, \qquad [\eta^{-1}] = 
\mass\cdot\length\cdot\timeunit^{-1}.
\end{equation}
Note that similar to the electrodynamics~\cite{DPRZ2017}, the inverse product $ 
(\epsilon\,\mu)^{-1} $ scales as the velocity square
\begin{equation}
\frac{1}{\epsilon\,\mu} \sim \Vel^2.
\end{equation}

\section{Turbulence as a non-equilibrium state}\label{sec.recovering.NS}

In this section, we discuss a non-equilibrium thermodynamics point of view on the emergence 
of turbulence as an \textit{excitation} of the laminar flow, which is treated as a \textit{near 
equilibrium} state 
of the fluid in a certain sense specified in what follows. In the discussion, we 
shall 
distinguish three length scales: the scale of interest (the problem length scale) $ \Lambda $, 
the 
macroscopic observer length scale $ L $ which we identify with the resolution scale $ \Delta x $ 
(see Section\,\ref{sec.observer}), 
and the viscosity dominated microscale which can be identified with the so-called Taylor 
microscale. The 
latter has been denoted by $ \ell $ in Section\,\ref{sec.basic}. In particular, the extended 
model~\eqref{PDE.extend} is essentially formulated in the settings when $ L=\Delta x > \ell $, 
that 
is we are not in the direct numerical simulation (DNS) conditions (recall that $ \Re \sim L/\ell 
$, 
see \eqref{def.Re}). Therefore it is implied that 
\begin{equation}
\ell < L < \Lambda.
\end{equation}

We say that, for a given macroscopic observer $ 
L= \Delta x $,  the flow is \textit{turbulent} or \textit{non-equilibrium} if at this scale, the 
following \textit{local} equilibrium 
condition is violated. Otherwise, the flow is treated as \textit{near equilibrium} or 
\textit{laminar}. Thus, the extended model~\eqref{PDE.extend} has equilibrium 
states of two types:
\vspace{-3mm}
\begin{enumerate}
\item \textit{Global equilibrium state}, which is the state when all thermodynamic forces 
vanish (it corresponds 
to the Euler equations of ideal fluids): 
\begin{equation}\label{equil.glob}
\calE_{\Dist} = 0, \qquad \calE_{\Burg} = 0, \qquad \calE_{\Durg} = 0 .
\end{equation}
\item \textit{Local equilibrium state}, which is a state when the thermodynamic forces do not 
identically
vanish but 
the 
non-equilibrium source term in~\eqref{PDE.extend.D} vanishes (see also 
Remark~\ref{rem.relax.limit}):
\begin{equation}\label{equil.loc}
\eta^{-1}\calE_\Durg - \alpha^{-1}\calE_\Dist = 0.
\end{equation}
\end{enumerate}

In other words, in the vicinity of local equilibrium state, i.e. when the left 
hand-side of \eqref{equil.loc} is small, we say that the flow is laminar at the scale $ L =\Delta 
x 
$ even though it may look highly irregular at the scale of interest $ \Lambda $ (the largest 
scale). In this 
case, the flow can be 
well approximated by the basic model~\eqref{PDE.basic} or equivalently, by the Navier-Stokes 
equations. Moreover, we shall explicitly demonstrate that, in this terminology, the 
DNS conditions, $ L = \ell $ 
or $ \Re = 1 $, correspond to near-equilibrium conditions (small left hand-side in 
\eqref{equil.loc}). 
But 
first, an important remark is in order.

\begin{remark}
The condition \eqref{equil.loc} is scale dependent 
because the relaxation parameter $ \alpha $ scales as inverse length $ \alpha \sim 
\length^{-1}$. It is 
therefore not surprisingly that the condition \eqref{equil.loc} is adjustable. For example, if we 
identify $ \alpha $ with the macroscopic observer length scale $ \Delta x^{-1} $ then, for an 
observer $ \Delta x' $, exactly the same flow may look ``more equilibrium'' (more ``laminar'') 
than 
for another 
observer
$ 
\Delta x'' > 
\Delta x' $. In other words, the extent to what the flow can be regarded as equilibrium 
(laminar-like) or non-equilibrium (turbulent-like) is the matter of at what scale we observe the 
flow (e.g. 
it is the matter of resolution of the computational mesh).
\end{remark}

%Thus, we treat a flow as laminar if, 
%at 
%the observation length scale $ L $ (which we have identified with the computational cell size $ 
%\Delta x $ in Section\,\ref{sec.observer}), it is well approximated by the 
%solution of the basic system~\eqref{PDE.basic} \revOne{Q5.}{or equivalently, by the solution of 
%the 
%Navier-Stokes equations}. Such a solution, as explained below, is identified 
%with the \textit{local equilibrium state} (see condition \eqref{equil.loc}) of the 
%non-equilibrium 
%extended 
%model~\eqref{PDE.extend}. \revOne{Q5.}{For example, according to this definition, any solution 
%obtained by means of the direct numerical simulation (DNS) is laminar because at the macroscopic 
%observer length scale (i.e. inside of each computational cell), the flow is well described by the 
%steady-state Newtons' law of viscosity and, by definition, does not contain any transient 
%structures such as vortices.}

We now note that $ \eta \sim \theta^{-1} \alpha^2 $ (their physical units are the same, see 
Section\,\ref{sec.units}), and it is 
convenient to assume that
\begin{equation}
\eta = \theta^{-1} \alpha^2.
\end{equation}
Furthermore, we define the new relaxation parameter $ \lambda = \alpha / \theta$ which, after
identifying the scaling parameter $ \alpha $ with the inverse length scale of the macroscopic 
observer $ L = \Delta x $, scales as Reynolds number~\eqref{def.Re}:
\begin{equation}\label{def.lambda}
\lambda = \frac{\alpha}{\theta} \sim \frac{L^{-1}}{\rho \tau \csh^2} \sim 
\frac{T}{\mass}\cdot\frac{L}{\ell} \approx \frac{L}{\ell} = \Re,
\end{equation} 
where $ T $ is the time scale of the macroscopic observer.

Hence, the source term in~\eqref{PDE.extend.D} can be rewritten as
\begin{equation}
-\frac{\theta}{\alpha} \left (\alpha^{-1}\calE_{\Durg} - \theta^{-1}\calE_{\Dist}\right ) \sim 
-\frac{1}{\Re} \left (\alpha^{-1}\calE_{\Durg} - \theta^{-1}\calE_{\Dist}\right ).
\end{equation}
Therefore, in the relaxation limit $ \lambda 
\rightarrow 0 $ ($ \Re \rightarrow 0  $), i.e. when approaching local 
equilibrium~\eqref{equil.loc}, the basic system~\eqref{PDE.basic} is recovered (see 
Remark\,\ref{rem.relax.limit}) which does contain the 
Navier-Stokes solutions as a particular case~\cite{DPRZ2016}. 
In particular, the DNS conditions, $ \Re \sim L/\ell \leq 1 $, correspond to the 
near-equilibrium settings and hence, both the basic 
model~\eqref{PDE.basic} and 
the Navier-Stokes model can be still used to describe turbulent flows under the DNS conditions.
On the other hand, if $ \Re 
\rightarrow \infty $, or equivalently $ L \gg \ell $, the local equilibrium 
condition~\eqref{equil.loc} is not 
respected in 
general, and the basic model, as well as the Navier-Stokes one, is not applicable.

One may note that in this framework, the turbulence is a non-linear interplay of the irreversible 
and 
dispersive 
processes governed by the thermodynamic forces $ \calE_{\Dist} $ and $ \calE_{\Durg} 
$ and by the new relaxation parameter $ \lambda $, which is essentially the Reynolds number of the 
macroscopic observer $ L = \Delta x $.

\section{Angular momentum conservation}\label{sec.ang.mom}

There are two aspects regarding the angular momentum conservation. The first one is 
due to the choice of the local (point-dependent) observer and related to such an observer 
\textit{spurious} 
Coriolis forces. This is, however, not about the material modeling but about the proper choice of 
the spacetime geometry and can be addressed in a very similar fashion as it is done in the special 
relativity via the adoption of the spin connection (or inertial connection) similar to the one in 
\eqref{spin.connection}, e.g. see \cite{AldrovandiPereiraBook}. Such a covariant theory will 
take into account the spurious inertial forces related to the local observer by construction 
(including the conservation of angular momentum). In this paper, 
however, we are not 
considering such a fully covariant extension and planning to do it in the future.

%In a polar medium, the balance of microscopic angular momentum reads 
%\cite{Eringen1968,Steinmann1994,Ehlers2018}
%\begin{equation}\label{angular.mom}
%\rho\frac{\rmd s_i}{\rmd t} - \frac{\pd m_{ki}}{\pd x^k } = \LeviCivitaUp{ikl}T_{kl} + \rho c_i,
%\end{equation}
%where $ \rmd/\rmd t $ is the material time derivative, $ s_i = \Theta_{ij}\omega_j$ is the spin, $ 
%\Theta_{ij} $ is the inertia tensor 
%of 
%microrotation per unit mass, $ \omega_i $ is the microrotation associated angular velocity, $ 
%m_{ik} $ is the couple stress tensor, $ c_i $ is the vector of body couple. Equation 
%\eqref{angular.mom}, in particular, says that if the body couples are absent $ c_i = 0 $ and the 
%stress tensor is symmetric $ T_{ik} = T_{ki} $ then the angular momentum is conserved.

The second aspect is due to the presence of the microstructure with a certain 
characteristic length scale associated with $ \alpha^{-1} $ in Section\,\ref{sec.recovering.NS}.
Under many circumstances, the presence of microstructure results in non-negligible inertial 
effects (microinertia) which have to be taken into account in the macroscopic balance equations 
(balance of linear and angular momentum).

%The non-vanishing microstructure length scale results in the presence of microinertia, 
%couple 
%stress and non-symmetric Cauchy stress. Such media are called micropolar or Cosserat 
%media~\cite{Ehlers2018}. 

Usually, in continuous theories\footnote{For example, see 
\cite{Toupin1962,Dahler1963,Eringen1968,Steinmann1994}} of media with a microstructure that posses 
rotational degrees of freedom (micropolar 
continua), the balance of linear momentum is written only for the macroscopic momentum (the average 
of the center-of-mass momenta of the microstructure elements) $ \mm = \rho\vv $:
\begin{equation}\label{momentum.mech}
\frac{\pd m_i}{\pd t} + \frac{\pd \left (v^k m_i - \sigma^k_{\ i}\right )}{\pd x^k}  = 0.
\end{equation}
This implies that the momentum flux $ t^k_{\ i} = v^k m_i - \sigma^k_{\ \, i} $ is not symmetric 
(equivalently, $ \sigma^k_{\ i} $ is not symmetric due to the symmetry of $ v^k m_i = \rho v^k 
v_i$) in general 
because \eqref{momentum.mech} does not account for the inertial forces 
experienced by the elements of the microstructure due to their micropolar rotations.

In such theories, in order to account for the inertial effects due to the micropolar 
rotations, the 
concept of the couple stress is introduced which emerges as the constitutive flux in the 
balance of micropolar angular momentum, e.g. see \cite{Eringen1968,Steinmann1994,Ehlers2018}.

On the other hand, the balance of linear momentum \eqref{PDE.extend.M} of our theory 
\begin{equation}\label{momentum.total}
\frac{\pd M_i}{\pd t} + \frac{\pd \left (v^k M_i - \Sigma^k_{\ i}\right )}{\pd x^k}  = 0
\end{equation}
is the conservation law of the \textit{total} momentum $ \MM = \mm + \ldots $ of the system which 
also includes 
the momentum of the torsion fields associated with the microstructure, see 
\eqref{momentum.total.example}. Therefore, because 
the total angular momentum, that is the sum of the angular momenta of all the constituents of the 
system including any fields, is a fundamentally conserved quantity, the total momentum flux $ 
\MomFlux{k}{i} = 
v^k M_i - \Cauchystr{k}{i} $ has to be symmetric. This is 
similar to the electrodynamics of moving medium where the sum of the  angular momentum of the 
medium and electromagnetic field is conserved \cite{Jackson1999}. Yet, we recall that the parts of 
the momentum flux, that is the convective part $ v^k M_i $ and the constitutive part $ 
\Cauchystr{k}{i} $, are not symmetric in general, see Section\,\ref{sec.closure}.

\section{Variational nature of the extended system and relation to micromorphic 
continua}\label{sec.Hamilton}

In this section, we demonstrate that the \textit{reversible part} (which does not raise the 
entropy), i.e. the left hand-side of 
\eqref{PDE.extend} and the reversible source terms can be obtained from 
the variational principle. 
Also, the variational formulation allows to clarify the physical meaning of the new state 
variables. Thus, from what follows, it becomes clear that $ \Burg $ represents the non-local 
interaction in space while $ \Durg $ represents non-locality in time and can be associated to 
microinertia.

\subsection{The case of holonomic triad}

We first need to recall the standard variational formulation of the continuum mechanics in order 
to 
see how it should be generalized for the case of anholonomic triads. In the following, we shall 
demonstrate how to derive the left hand-side of the basic model~\eqref{PDE.basic} from the 
variational principle.
 
Similar to the GENERIC formulation of non-equilibrium 
thermodynamics~\cite{PKG-Book2018,Ottinger2005}, in the SHTC framework, the reversible and 
irreversible parts of the time evolution are treated separately. Thus, the reversible part of 
\eqref{PDE.basic} is 
Hamiltonian and can be obtained either from Hamilton's principle of stationary action or can be 
generated by corresponding Poisson brackets, e.g. see \cite{SHTC-GENERIC-CMAT}. The irreversible 
part is represented by local (algebraic) terms of relaxation type, $ -\frac{1}{\theta} E_{\Dist} 
$ and $ \frac{1}{\calE_s \theta}E_{\Dist}:E_{\Dist} $, and is derived from the second 
law of thermodynamics~\cite{SHTC-GENERIC-CMAT}. Hence, in the absence of dissipation, the 
deformations 
are reversible and the Lagrangian formalism can be utilized, that is the variational principle in 
the 
SHTC framework is formulated on the globally relaxed matter manifold\footnote{Note that, in 
principle, the 
equation of motion (the energy-momentum conservation) can be also obtained in the Eulerian 
coordinates. However, the time evolution of the triad field (deformation gradient in this case) 
can 
be only obtained in an \textit{ad hoc} manner, e.g. see \cite{Gavrilyuk2011}. While, if formulated 
in the Lagrangian coordinates, the triad evolution is derived 
rigorously~\cite{SHTC-GENERIC-CMAT}.} 
$ \MatterO $ parameterized with the Lagrangian coordinates $ \myxi{A} $, $ \sA = 1,2,3 $. Also, 
despite we are working in the non-relativistic settings and the time is absolute, it is convenient 
to distinguish notations for the time coordinate. Thus, we use the chart $ (\tau,\myxi{A}) $ for 
the 
four Lagrangian coordinates, while the chart $ (t,x^i) $ stands for the Eulerian (laboratory) 
coordinate system even 
though $ \tau = t $. For 
example, in this notations, the so-called material time derivative is $ \pd_\tau = \pd_t + 
v^i\pd_i $ with $ v^i $ being the material element velocity.

The existence of motion $ x^i = \motion{i}(\tau,\myxi{A}) $ 
connecting in an one-to-one manner the manifolds 
$ \MatterO $ and $ \Matter $
plays the central role because the Lagrangian density is defined as
\begin{equation}\label{def.Lagrangian.density}
\tilde{\Lambda}(\tau,\myxi{A},\motion{i},\pd_\tau\motion{i},\pd_{\sA} \motion{i}) = 
\Lambda(\pd_\tau\motion{i},\pd_{\sA}\motion{i}),
\end{equation}
where 
$ \pd_\tau:= \frac{\pd}{\pd \tau} = \pd_t + v^i \pd_i $, $ \pd_{\sA} := \frac{\pd}{\pd \myxi{A}} $
are the Lagrangian 
time and spatial derivatives. Variation of the action $ \int \Lambda \rmd \tau\rmd \bm{\xi} $ with 
respect to the motion $ \motion{i} $ gives the Euler-Lagrange equations (momentum conservation)
\begin{equation}\label{eqn.EL.1}
\pd_\tau \Lambda_{v^i} + \pd_{\sA} \Lambda_{\defgrad{i}{A}} = 0,
\end{equation}
where the notations
\begin{equation}\label{def.vel.def.grad}
 v^i := \pd_\tau \motion{i}, \qquad \defgrad{i}{A} := \pd_{\sA}\motion{i} ,
\end{equation} 
were introduced and have the meaning of the velocity and the deformation gradient respectively. 
The 
Euler-Lagrange equations~\eqref{eqn.EL.1} are under-determined equations (we have 12 unknowns $ 
v^i 
$ and $ 
\defgrad{i}{A} $ and only three equations) and thus, have to be supplemented with nine more 
evolution equations. These equations are given by the time evolution of the deformation gradient
\begin{equation}\label{def.grad.evol}
\pd_\tau \defgrad{i}{A} - \pd_{\sA}v^i = 0,
\end{equation} 
which are the trivial consequences of the definitions \eqref{def.vel.def.grad} and thus, they play 
the role of the integrability conditions for the Euler-Lagrange equations~\eqref{eqn.EL.1}. 
Equations \eqref{eqn.EL.1} and \eqref{def.vel.def.grad} together form a closed system of twelve 
evolution 
PDEs for twelve unknowns $ (v^i,\defgrad{i}{A}) $ if only the Lagrangian density $ 
\Lambda(v^i,\defgrad{i}{A}) $ is specified (closure). These are however the equations written in 
the Lagrangian coordinates $ (\tau,\myxi{A}) $ and have to be transformed into the Eulerian 
coordinates in order to obtain left hand-side of the equations~\eqref{PDE.basic.m} and 
\eqref{PDE.basic.A}, see details in~\cite{SHTC-GENERIC-CMAT}.

\subsection{The case of anholonomic triad}

The variational scheme described in the previous section is applicable only to the case when the 
triad $ 
\defgrad{i}{A} $ is holonomic, i.e. it is the gradient of the single-valued map $ 
\motion{i}(\tau,\myxi{A}) $. Because our main intention is to deal with irreversible deformations 
of fluids (or solids), the original globally relaxed configuration of the continuum $ \MatterO $ 
plays no 
role in the evolution of flowing matter. Instead, a new locally relaxed configuration $ \MatterR 
$ was 
introduced in Section\,\ref{sec.diffgeom}. As was discussed there, such a locally relaxed manifold 
$ \MatterR $ can not be connected with $ \Matter $ in a one-to-one manner and hence, a map, like $ 
\motion{i}(\tau,\myxi{A}) $ between $ \MatterO $ and $ \Matter $, does not exist in general. In 
turn, this poses a serious problem in the classical way of the velocity field definition, that is 
via the time derivative of the motion \eqref{def.vel.def.grad}.

In fact, our Riemann-Cartan formulation of fluid dynamics admits a variational formulation which 
does not require the motion $ 
\motion{i}(\tau,\myxi{A}) $. This can be naturally explained in the four-dimensional formalism (we 
still stay in the non-relativistic settings)
because the velocity field is, in fact, a part of the anholonomic tetrad (4-distortion). Indeed, 
in the four-dimensional formalism, the classical definition of the velocity 
field~\eqref{def.vel.def.grad} can be viewed as a part of the 4-deformation gradient. Thus, let us 
consider the extended Eulerian coordinates $ x^\mu = (t,x^i) $, $ \mu = 0,1,2,3 $ and the 
Lagrangian 
coordinates $ \myxi{A} = (\tau,\xi^1,\xi^2,\xi^3)$, $ \sA = 0,1,2,3 $ which now include also the 
time coordinate. Yet recall that, in the non-relativistic settings, the time is assumed to be 
absolute, i.e. $ t = \tau $.
Hence, the velocity field $ v^\mu = \frac{\pd \chi^\mu}{\pd \xi^0} = (1,v^1,v^2,v^3)$ is just a 
part of the 4-deformation gradient $ \defgrad{\mu}{A} = \frac{\pd 
\motion{\mu}}{\pd \myxi{A}}$
\begin{equation}
\defgrad{\mu}{A} := 
\left( \begin{array}{cccc}
	       1         &    0     &    0     &    0     \\
	v^1 & F^1_{\ 1} & F^1_{\ 2} & F^1_{\ 3} \\
	v^2 & F^2_{\ 1} & F^2_{\ 2} & F^2_{\ 3} \\
	v^3 & F^3_{\ 1} & F^3_{\ 2} & F^3_{\ 3}
\end{array} 
\right).
\end{equation}

Motivated by this fact, we thus consider the velocity $ v^\mu $ and the anholonomic triad field $ 
\itriad{i}{a} $, $ a=1,2,3 $ (inverse of $ \triad{a}{i} $) as being parts of the anholonomic 
tetrad 
$ 
\itetrad{\mu}{a} $, $ a=0,1,2,3 $. In other words, the velocity field will be not defined as the 
time derivative of a motion but it is postulated axiomatically and is defined as the solution to a 
certain evolution equation, namely the momentum conservation law. 

The Lagrangian density is then considered as a function of the 
tetrad $ \itetrad{\mu}{a} $ and its first-order derivatives $ \pd_b\itetrad{\mu}{a} $
\begin{equation}\label{def.Lagr.density4}
\tilde{\Lambda}(\itetrad{\mu}{a},\pd_b\itetrad{\mu}{a}) = \Lambda(\itetrad{\mu}{a},\dualtors{\mu 
ab}),
\end{equation}
where $ \dualtors{\mu ab} :=\LeviCivitaUp{abcd}\pd_c \itetrad{\mu}{d} = \frac12 
\LeviCivitaUp{abcd} 
T^\mu_{\ 
ab}$ is the Hodge dual of the torsion $ T^\mu_{\ ab} := \pd_a\itetrad{\mu}{b} - 
\pd_b\itetrad{\mu}{a}$, and $ \LeviCivitaUp{abcd} $ is the 4D Levi-Civita symbol. Recall that we 
do 
not 
need 
to introduce a coordinate system in $ \MatterR $ in order to define the derivatives $ \pd_a $. 
Instead, we use that fact that the tangent spaces to $ \MatterR $  are soldered to $ \Matter $ and 
the material derivatives $ \pd_a $ are treated as the space derivatives $ \pd_\mu $ written in the 
local Cartan 
frame $ \bm{e}^a $, that is $ \pd_a := \itetrad{\mu}{a} \pd_\mu $, see \eqref{deriv.transform}.
Recall that, due to fixing gauge degrees of freedom (local rotations $ \Rot{a}{a}(x^i) $ in the 
tangent 
spaces $ \TMlr $), the derivatives $ \pd_a $ are, in fact, the covariant derivatives in $ \MatterR 
$, see the discussion right after equation~\eqref{deriv.transform}.

The 
first variation of the action gives us the Euler-Lagrange equations
\begin{equation}\label{PDE.4torsion.EL}
\pd_d(\LeviCivitaUp{abcd} \Lambda_{\dualtors{\mu bc}}) = \Lambda_{\itetrad{\mu}{a}},
\end{equation}
which have to be supplemented by the integrability conditions
\begin{equation}\label{PDE.4torsion.integrab}
\pd_b\dualtors{\mu ab} = 0.
\end{equation} 

So far, we obtain only equations for the torsion field $ T^\mu_{\ ab} $ which give us equations 
\eqref{PDE.extend.B} and \eqref{PDE.extend.D}, while we still do not have evolution PDEs for the 
tetrad field $ \itetrad{\mu}{a} $. In fact, the tetrad evolution PDE 
\begin{equation}\label{energy.momentum.4d}
\pd_a\Lambda_{\itetrad{\mu}{a}} = 0
\end{equation}
is exactly the momentum conservation law (the energy-momentum conservation to be more precise). 
This PDE, however, 
is not 
obtained by means of variation of the action like in the classical mechanics but it is an identity 
which is obtained after applying 
the divergence operator to \eqref{PDE.4torsion.EL} (the left hand-side then identically vanishes). 
Remark that such a variation scheme, though formulated in the matter manifold, is fully equivalent 
to the 
variation 
scheme in the teleparallel gravity formulated in the spacetime manifold, e.g. see~\cite{Cai2016}.

Because the targeting reader, we believe, is used to work in the three-dimensional formalism 
instead 
of the four-dimensional one described above, we shall rewrite now this variational formulation in 
the 3D notations.

%The variational 
%formulation of the SHTC equations is performed in the Lagrangian coordinates 
%because, in the Eulerian coordinates, the evolution equation for the triad $ \ $, the variational 
%principle 
%has to be formulated in the Lagrangian 
%coordinates  After the Euler-Lagrange equations are obtained they can be 
%transformed 
%in 
%to the Eulerian coordinates $ x^i $~\cite{SHTC-GENERIC-CMAT,DPRZ2017}. Thus, analogous to the 
%electromagnetic vector potential and scalar potential we introduce three vector potentials ($ 
%i=1,2,3 $) and 
%three 
%scalar potentials\footnote{Note that the velocity vector $ v^i $ is not actually a vector in the 
%Lagrangian space but rather three scalars because the components $ v^i $ transforms as scalars 
%with 
%respect to the change of Lagrangian coordinates $ \xi^a $.} (it is convenient to distinguish the 
%upper and lower indices in this section)
%\begin{equation}\label{def.Ham.potentials}
%\itriad{i}{a}(t,\xi^{a}), \qquad v^i(t,\xi^a),
%\end{equation}
%where $ \itriad{i}{a} $ is the inverse distortion field\footnote{In the solid dynamics 
%literature, 
%the inverse distortion $ \AA^{-1} $ is usually called the \textit{elastic deformation gradient} 
%and 
%denoted by $ 
%\bm{F}^e $, e.g. see~\cite{Hyper-Hypo2018}, while $ \bm
%F $ usually stands for the total deformation gradient.}, i.e. $ \triad{a}{i}\itriad{i}{b} = 
%\delta^a_{\ b}$ and $ \triad{a}{i} \itriad{j}{a} = \delta^i_{\ j}$, while $ v^i $ is the velocity 
%field.

Thus, we now treat the Lagrangian density \eqref{def.Lagr.density4} as the function of the 
potentials $ v^i $ and $ \itriad{i}{a} $ and their first derivatives (in what follows $ a,b,c = 
1,2,3 $ 
again) 
\begin{equation}\label{def.Ham.Lagr}
L(v^i,\itriad{i}{a},e^{i}_{\ 
a},h^{ai}),
\end{equation}
where \begin{equation}\label{def.Ham.newvar}
e^i_{\ a}:= \pd_\tau \itriad{i}{a} - \pd_a v^i, \qquad h^{ai} 
:=-\LeviCivitaUp{abc}\pd_b\itriad{i}{c},
\end{equation}
where, in turn, $ \pd_\tau $ is the material time derivative
and 
$ \pd_a = \itriad{i}{a} \pd_i $, $ a=1,2,3 $ are the material space derivatives. Note that the PDE 
\eqref{def.Ham.newvar}$ _1 $ looks as the PDE \eqref{def.grad.evol} for 
the total deformation gradient (which is holonomic triad) in the Lagrangian coordinates apart that 
\eqref{def.Ham.newvar}$ _1 $ 
has the \textit{source of non-holonomy} $ e^i_{\ a} $. 

Then, the first variation of the action with respect to $ v^i $ and $ \itriad{i}{a} $ gives the 
Euler-Lagrange equations:
\begin{equation}\label{PDE.Euler.Lagrange}
\pd_\tau L_{e^i_{\ a}} + \LeviCivitaUp{abc}\pd_b L_{h^{c i}} = L_{\itriad{i}{a}}, \qquad 
-\pd_a L_{e^i_{\ a}} = L_{v^i}.
\end{equation}
Exactly as for \eqref{energy.momentum.4d}, the momentum conservation (compare with 
\eqref{eqn.EL.1})
\begin{equation}\label{energy.momentum.3d}
\pd_\tau L_{v^i} + \pd_a L_{\itriad{i}{a}} = 0
\end{equation}
is obtained not by means of variations but it is the identity which is obtained by applying $ 
\pd_\tau $ to 
\eqref{PDE.Euler.Lagrange}$ _2 $ and $ \pd_a $ to \eqref{PDE.Euler.Lagrange}$ _1 $ and summing up 
the results.

Equations of motion $ \eqref{PDE.Euler.Lagrange} $ and \eqref{energy.momentum.3d} have to be 
supplemented with the integrability conditions (they are 
trivial consequences of the definitions~\eqref{def.Ham.newvar}):
\begin{equation}\label{PDE.Ham.integrab}
\pd_\tau h^{a i} + \LeviCivitaUp{abc}\pd_b e^i_{\ c} = 0, \qquad \pd_b h^{b i} = 0, \qquad 
\pd_\tau 
\itriad{i}{a} - \pd_a v^i = e^i_{\ a}.
\end{equation}
After the introduction of the new variables
\begin{equation}
m_i = L_{v^i}, \qquad \durgL{a}{i} := L_{e^i_{\ a}}, \qquad \burgL{a}{i} := h^{ai},
\end{equation}
and new potential 
\begin{equation}\label{LegendreUL}
U := \vel{i} L_{v^i} + e^i_{\ a} L_{e^i_{\ a}} - L
\end{equation}
as a partial Legendre 
transformation of $ L $, the equations \eqref{PDE.Euler.Lagrange}$ _1
$ and \eqref{PDE.Ham.integrab}$ _1 $ can be rewritten as (note that $ U_{\burgL{a}{i}} = 
-L_{\burgL{a}{i}} $, $ U_{\itriad{i}{a}} = -L_{\itriad{i}{a}} $)
\begin{subequations}
\begin{equation}\label{PDE.mA.Lagr}
\pd_\tau m_i - \pd_a U_{\itriad{i}{a}} = 0, \qquad
\pd_\tau \itriad{i}{a} - \pd_a U_{m_i} = U_{D^a_{\ i}},
\end{equation}
\begin{equation}\label{PDE.DB.Lagr}
\pd_\tau \durgL{a}{i} - \LeviCivitaUp{abc}\pd_b U_{\burgL{c}{i}} =-U_{\itriad{i}{a}}, \qquad
\pd_\tau \burgL{a}{i} + \LeviCivitaUp{abc}\pd_b U_{\durgL{c}{i}} = 0.
\end{equation}
\end{subequations}
These are the equations written in the relaxed matter manifold $ \MatterR $. Their 
structure may remind the structure of the electrodynamics equations studied 
in~\cite{DPRZ2017,SHTC-GENERIC-CMAT}. In particular,  one can use the series of cumbersome
coordinate, state variable and thermodynamic potential transformations from 
\cite{SHTC-GENERIC-CMAT} in order to transform these equations into their Eulerian 
counterparts~\eqref{PDE.extend}.
Nevertheless, we remark that the Hamiltonian structure of \eqref{PDE.DB.Lagr} (i.e. 
the underlined Poisson brackets) is not fully equivalent to the Hamiltonian structure of the 
electrodynamics equations due to the presence of the reversible source terms, see 
Section\,\ref{sec.extended} which are the part of the reversible part of the time evolution and 
therefore have to be generated by the Poisson brackets.
Remark that the energy potential $ \calE $ in \eqref{PDE.extend} and the 
potential $ U $ in \eqref{PDE.DB.Lagr} are related as $ \calE = w \, U $ if $ w = \det(\Dist) $ 
\cite{SHTC-GENERIC-CMAT}, and hence $ \calE $ is directly related to the Lagrangian $ L $ via the 
Legendre transformation \eqref{LegendreUL}.

%Recall that in the SHTC framework, the reversible part of the time evolution (which conserves 
%both 
%the entropy and the total energy) is Hamiltonian, while the irreversible part (which conserves 
%the 
%energy but raises the entropy) can not be obtained from Hamilton's principle and therefore is 
%treated separately. The irreversible part of the time evolution is constructed
%based on the laws of thermodynamics, see~\cite{SHTC-GENERIC-CMAT}. That is why the counterpart to 
%the 
%dissipative source term in \eqref{PDE.extend.D} is missing on the right hand-side of
%\eqref{PDE.DB.Lagr}$ _1 $.

Finally, we note that the variational viewpoint on the extended model~\eqref{PDE.extend} allows to 
see 
more 
clearly the physical meaning of the new fields $ \Durg $ and $ \Burg $. Thus, from the definition 
\eqref{def.Ham.newvar}$ _1 $ it becomes clear that the 
complimentary filed $ \Durg $ represents the \textit{non-locality in time} and can be associated 
to 
the micro-inertia of the microstructure (either emerging as in turbulence or preset from the 
beginning as in the 
micro-structured solids). In particular, such an observation can be useful for modeling of 
microscopically heterogeneous materials which are called acoustic metamaterials and/or phononic 
crystals, which, in the last years, raised  a lot of interest in mathematical modeling in solid 
mechanics and show exotic response with respect to perturbations of certain wave 
lengths~\cite{Cummer2016}. To model such materials, higher gradients (in space and time) of a 
total and microscopic deformation field are adopted. A quite general class of such models, called 
\textit{micromorphic} or \textit{generalized continua} 
models, has 
been 
developed by many authors starting from the work by Cosserat brothers~\cite{Cosserat1909}, 
Mindlin~\cite{Mindlin1964} and Eringen~\cite{Eringen1968}, in the 
context of linear deformations, e.g.
see~\cite{Forest2013,LeMarec2014,Madeo2016,Barbagallo2017,DellIsola2018a} and the references 
therein, and finite deformations~\cite{Berezovski2011a,Berezovski2011,Bohmer2016,Bohmer2018}. The 
proposed 
\textit{non-linear} 
model~\eqref{PDE.extend} can be also attributed to the class of micromorphic models since the 
higher spatial and time gradients of the local distortion field are used as independent state 
variables. In particular, we expect that the non-linear dispersive models like the one proposed in 
this paper can 
be applied 
to modeling of 
genuinely nonlinear waves such as solitons and shock waves in microstructured solids.

\section{Galilean and Lorentz invariance}

In Newtonian mechanics, we used to deal with material equations which have to respect the Galilean 
invariance principle. The basic equations~\eqref{PDE.basic} as well as other equations belonging 
to 
the SHTC class such as hyperbolic mass and heat conduction~\cite{SHTC-GENERIC-CMAT} are Galilean 
invariant by construction~\cite{Godunov1996}. This is due to the Hamiltonian nature of the SHTC 
equations. 

The extended system~\eqref{PDE.extend} is however not a true material system in the sense that the 
new fields $ \Burg $ and $ \Durg $ are like the classical fields such as electric and magnetic 
fields, for example, that do not have a material carrier. Moreover, as we could see, there is a 
straightforward analogy between the structure of the Maxwell equations in moving 
medium~\cite{DPRZ2017} and the 
extended model~\eqref{PDE.extend}. It also immediately follows from this analogy that the 
equations 
for $ \Burg $ and $ \Durg $ fields are not Galilean invariant but Lorentz invariant. For example, 
the characteristic analysis in~\cite{DPRZ2017} shows that the maximum sound speed in such a medium 
is $ c = 1/\sqrt{\epsilon\mu} $ instead of one would expect $ v \pm c $ which is inherent 
to Galilean invariant systems, where $ v $ denotes the medium velocity in the given coordinate 
direction.

\section{Conclusion and perspectives}

We have presented an extension of the unified hyperbolic formulation of continuum fluid and solid 
mechanics recently proposed in~\cite{HPR2016,DPRZ2016,HYP2016}  mentioned above as the basic 
system~\eqref{PDE.basic}. The basic system, in turn, is based on the Godunov-Romenski model for 
elastoplastic deformation in metals developed in 1970s~\cite{GodRom1974,God1978,Romenski1979}.
The extended system aims in accounting for the rotational degrees of freedom of the main field of 
the basic model, the distortion field, denoted as $ \Dist $. 

In Section\,\ref{sec.diffgeom}, the distortion field is considered as the Cartan moving frame and 
thus it represents anholonomic basis triad field
in general. As the characteristic of anholonomy, we use the torsion tensor. The distortion and 
torsion are therefore define completely the internal non-Euclidean (Riemann-Cartan) flow geometry 
with \Weitz connection as a linear affine connection. 
The paper is 
therefore can be treated as an attempt to introduce methods of differential geometry to fluid 
dynamics in general and turbulence modeling in particular. Moreover, the proposed theory is 
essentially an analog of the teleparallel 
gravity (torsional 
gravity)~\cite{KleinertMultivalued,AldrovandiPereiraBook,Cai2016,Golovnev2017a}, 
which can be viewed 
as a Yang-Mills-type translational gauge theory~\cite{Hayashi1967,Cho1976,Hayashi1977,Lazar2000}. 
It is 
therefore would be also 
interesting to consider the turbulence phenomena from the standpoint of a gauge field theory.
In this paper, we however have considered a version of the model with a fixed gauge. Namely, we 
assumed 
that the anholonomic frames in the relaxed tangent space are all co-aligned. The general case, 
without a gauge fixing and hence, 
with 
non-zero spin connection will be considered elsewhere.

In the fluids dynamics context, the anholonomy of the distortion field is mainly due to its 
rotational degrees of freedom (spin) which can be used to characterize the kinetic energy of the 
unresolved 
eddies in a turbulent flow. The torsion field $ \Burg $ is therefore was used to account for the 
distortion spin in the model and thus, $ \Burg $ plays the role of an independent state 
variable in the extended model along side with the 
complimentary field $ \Durg $. The former has 
the meaning of the number density of unresolved eddies while the latter has the meaning similar to 
the angular momentum and characterizes micro inertia as discussed in Section\,\ref{sec.Hamilton}. 
We then observed that in 
the extended model, the contribution due 
to torsion $ \Burg $ and $ \Durg $ field affects the 
definitions of the total momentum density, stress tensor, total energy and the flux of the total 
energy.

Furthermore, in order to fulfill the thermodynamic consistency, the dispersive terms (which 
conserve the entropy) have had to be introduced. These terms are of algebraic relaxation type and 
do not affect hyperbolicity of the model.

In Section\,\ref{sec.recovering.NS}, we identify the laminar-to-turbulent transition as 
the excitation of the laminar state,
which, in turn, is considered as a local equilibrium state. This transition is highly non-linear 
and is a 
result of the interplay between irreversible and dispersive mechanisms in the new model. The new
relaxation parameter~\eqref{def.lambda} governing such a transition is essentially the Reynolds 
number of the macroscopic observer.

We then observed that the governing equations for the new 
fields $ \Burg $ and $ \Durg $ share some elements of the common structure with the  
equations for electrodynamics of 
moving 
media recently 
studied in~\cite{DPRZ2017,SHTC-GENERIC-CMAT}. Also, similar to the electrodynamics, the fields $ 
\Burg $ and $ \Durg $ like the magnetic and electric fields are true classical fields because 
their 
carriers are mass-less entities like vortexes in fluids or defects in solids. In the solid 
dynamics 
context, the straightforward 
analogy between the equations governing the dislocation dynamics and those governing the 
electromagnetic fields was observed by several authors, 
e.g.~\cite{Kosevich1965,KadicEdelen1983,Grinyaev2000,Lazar2002a}. Also, similar to the 
electrodynamics equations, the momentum density $ \MM $ is treated as the total momentum (matter 
momentum $ + $ field's momentum). Due to this fact, the conservation of the total angular momentum 
is guaranteed by the symmetry of the full momentum flux tensor and thus, no additional equation for 
the balance of angular momentum is required in our theory, see Section\,\ref{sec.ang.mom}.

In Section\,\ref{sec.Hamilton}, we discussed the variational nature of the governing 
equations~\eqref{PDE.extend}. In 
the absence of dissipation, we formulated the variational principle without referencing to the 
initial globally relaxed configuration $ \MatterO $. Instead, only locally relaxed $ \MatterR $ 
and 
current $ \Matter $ configurations were used. This is an important result showing that the notion 
of motion encoding the full history of the medium evolution is not required for formulation of the 
variational principle.

Importantly, the new system is a nonlocal PDE system due to the fact that the torsion tensor is 
defined as higher gradients of the triad $ \Dist $. The nonlocality at small scales (both in space 
and time), i.e. when 
behavior at a point can be strongly influenced by the flow remote from that point, is 
within the main features of turbulent flows and is 
within the most difficult features to be modeled~\cite{Pope2001,Tsinober2009a}.  
Also recall that the torsion is 
specifically
used to describe nonlocal interaction (action at a distance) in tetrad formulations of 
relativistic theories such as teleparallel theories of gravity. One could therefore expect that 
the 
same 
geometrical concepts can be used to describe nonlocal effects in fluid dynamics. Besides, note 
that 
mathematical analogies between the non-linear electromagnetism and tetrad (or torsion) theories of 
gravity 
similar to those observed in this paper was recently discussed in~\cite{Hohmann2018a}.

So far, we did not discuss possible numerical implementation issues of the proposed model. We 
expect that solving the extended model numerically is not a trivial task due to the presence of
\textit{stiff relaxation terms} of both natures, irreversible and dispersive, and due to the 
strong 
non-linearity of the model. However, despite the strong non-linearity of the extended model, one 
can guaranty the local well-posedness (hyperbolicity) of the Cauchy problem if the energy 
potential 
is convex~\cite{SHTC-GENERIC-CMAT}.
Nevertheless, one may expect that the standard numerical methods for hyperbolic PDEs can be used 
only exceptionally but 
methods 
with the \textit{asymptotic preserving} and \textit{well-balanced} properties should be 
employed~\cite{DumbserEnauxToro,Castro2008,Russo2000}. 
Thus, as in the case of the basic model~\eqref{PDE.basic}, the family of ADER (Arbitrary 
high-order DERivatives) Finite Volume (ADER-FV) or Discontinuous Galerkin (ADER-DG) schemes, see 
\cite{DPRZ2016,DPRZ2017} and references therein, can successfully handle such issues in the 
asymptotic preserving and well-balanced manner. We therefore plan to study the extended model 
numerically and compare its solution against some experimental data as well as the results of 
Direct 
Numerical Simulation (DNS) of turbulent flows in subsequent papers.

One of the natural possibilities for the further extension of the present results 
might be obtaining of general/special relativistic versions of the basic and extended models. 
Recall that the classical Navier-Stokes fluid dynamics can not be applied to modeling relativistic 
flows due to the intrinsic causality and stability issues of the Navier-Stokes 
equations~\cite{RezzollaZanottiBook}. To fix these issues of the classical fluid dynamics, 
several alternatives have been developed 
\cite{Muller1967,MullerRuggeri1986,Israel1976,Stewart1977,Geroch1990,Freistuhler2017,Ottinger2019} 
to name just a few. The geometrical nature of our theory with the underlying Riemann-Cartan 
geometry also suggests that it would be not difficult to obtain the relativistic version of the 
theory which should fit well to the geometrical settings of the general relativity or its
teleparallel equivalent~\cite{AldrovandiPereiraBook}. This work is currently in progress.

The diversity of physical theories where the geometrical methods and concepts are used (e.g. 
gravity, differential forms in electromagnetism, Lagrangian and Hamiltonian mechanics, Yang-Mills 
gauge theory) may witness 
about their universal validity and independence on the physical context. We thus hope 
that this paper will help to reformulate modern Navier-Stokes-based fluid dynamics in a more 
universal language of differential geometry. This, in turn, may help to look at the turbulence 
problem 
from a fundamentally different angle.

\subsection*{Acknowledgement} 
The authors are thankful to A.\,Golovnev for the discussions of concepts of teleparallel 
gravity. Also, the authors acknowledge stimulating discussions with L. Margolin on the concept of macroscopic observer. 
Furthermore, 
I.P.
greatly acknowledges the support by ANR-11-LABX-0040-CIMI within the program ANR-11-IDEX-0002-02.
Results by E.R. obtained in Sections 5,\,7 were supported by RSF grant (project 19-77-20004).
The work by M.D. was partially funded by the European Union's Horizon 2020 Research and Innovation  Programme under the project \textit{ExaHyPE}, grant no. 671698 (call FETHPC-1-2014), Istituto Nazionale di Alta Matematica (INdAM) via the GNCS group and the program \textit{Young Researchers Funding 2018}. 
Furthermore, MD acknowledges funding from the Italian Ministry of Education, University and Research (MIUR) in the frame of the Departments of Excellence Initiative 2018--2022 attributed to DICAM of the University of Trento, as well as financial support from the University of Trento in the frame of the  \textit{Strategic Initiative Modeling and Simulation}. 
%The authors are also very grateful to the three anonymous referees and their constructive comments that were highly %appreciated and which helped to improve the quality and clarity of the paper.
% . 

\printbibliography

\end{document}